%% file: main.tex
\documentclass{LMCS}
\usepackage{enumerate}
\usepackage{epsfig}
\usepackage{graphicx}
\input xy
\xyoption{all}

\usepackage{clock}

\newcommand\forcecommand[1]{%
  \let#1\relax
  \newcommand#1%
}

\input{macros.tex}

\usepackage{amsmath,amssymb}

\theoremstyle{plain}

\theoremstyle{plain}

\theoremstyle{plain}\newtheorem{theorem}[thm]{Theorem}

\theoremstyle{plain}\newtheorem{proposition}[thm]{Proposition}

\theoremstyle{plain}

\theoremstyle{plain}

\title{ A Faithful Semantics for \\ Generalised Symbolic Trajectory Evaluation } 
\author[K.~Claessen]{Koen Claessen\rsuper a}
\address{{\lsuper a}Chalmers University of Technology, Sweden}
\email{koen@chalmers.se}
\thanks{}

\author[J.-W.~Roorda]{Jan-Willem Roorda\rsuper b}
\address{{\lsuper b}Fenix Design Automation, the Netherlands}
\email{janwillem@fenix-da.com}
\thanks{{lsuper b}This work was carried out while employed at Chalmers University}

\def\doi{5 (2:1) 2009}
\lmcsheading%
{\doi}
{1--32}
{}
{}
{Jun.~\phantom{0}7, 2007}
{Apr.~\phantom{0}8, 2009}
{}   

\begin{document}




\input{abstract.tex}

 
\maketitle\vfill\eject

\input{introduction.tex}

\input{ste.tex}

\input{gste-semantics.tex}

\input{gste-semantics2.tex}
\input{comparing.tex}

\input{gste-mc.tex}

\input{future.tex}

\input{conclusion.tex}

\bibliographystyle{plain}
\bibliography{main}


\end{document}

%% file: macros.tex
\forcecommand{\SeqGraph}{\mathrm{SeqGraph}} 
\forcecommand{\vstart}{\mathrm{start}}
\forcecommand{\vend}{\mathrm{end}}

\forcecommand{\Fsg}{F^{\circ}}
\forcecommand{\FsgTwo}{\Fsg_{\Sigma}}
\forcecommand{\FsgLine}{\Fsg_{\mathrm{line}}}
\forcecommand{\FsgNoCycle}{\Fsg_{\mathrm{nocycle}}}

\forcecommand{\Fseq}{F^{\rightarrow}}
\forcecommand{\ftF}{\Fseq}      

\forcecommand{\pre}{\mathrm{pre}} 

\forcecommand{\gfp}{\mathrm{gfp}}

\forcecommand{\Tau}{T}

\forcecommand{\ine}{\mathrm{in}}
\forcecommand{\oute}{\mathrm{out}}
\forcecommand{\init}{\mathsf{init}}

\forcecommand{\nextf}{\mathrm{next}}

\forcecommand{\ant}{\mathit{ant}}
\forcecommand{\cons}{\mathit{cons}}

\forcecommand{\gsequence}{sequence graph}
\forcecommand{\gtrajectory}{trajectory graph}
\forcecommand{\pseudodeftraj}{pseudo defining trajectory}
\forcecommand{\deftraj}{defining trajectory graph}
\forcecommand{\SG}{\mathrm{SeqG}}

\forcecommand{\AL}{\mathrm{AL}}

\forcecommand{\head}[1]{\noindent{\bf #1}$\;$}

\forcecommand{\comment}[1]{$\Longrightarrow${\bf\em{}#1}$\Longleftarrow$ \\}
\forcecommand{\todo}[1]{\comment{TODO: #1}}

\forcecommand{\weg}[1]{} 
\forcecommand{\watruimte}{\vspace{0.2cm}}
\forcecommand{\boxend}{\vspace{-0.7cm}\flushright{$\Box$}}
\forcecommand{\vboxend}{\vspace{-1.25cm}\flushright{$\Box$}}

\forcecommand{\ttt}{1}
\forcecommand{\fff}{0}

\forcecommand{\power}{\mathbb{P}}
\forcecommand{\twocases}[4]
{\left \{ \begin{array}{ll}#1, & #2 \\ #3, & #4 \end{array} \right.}
\forcecommand{\threecases}[6]
{\left \{ \begin{array}{ll}#1, & #2 \\ #3, & #4 \\ #5, & #6 \\ \end{array} \right.}

\forcecommand{\notg}{\mbox{\sc not}\;}
\forcecommand{\andg}{\;\mbox{\sc and}\;}
\forcecommand{\org}{\;\mbox{\sc or}\;}
 
\forcecommand{\band}{\mathrm{\wedge}}
\forcecommand{\bor}{\mathrm{\vee}}

\forcecommand{\sigmaS}{\sigma_{\State}}

\forcecommand{\sss}{\mathbb{P}(\States)}

\forcecommand{\sts}{\States \rightarrow \States}

\forcecommand{\X}{\mathsf{X}}
\forcecommand{\Top}{\mathsf{T}}
\forcecommand{\T}{\mathsf{T}}
\forcecommand{\glb}{\sqcap}
\forcecommand{\lub}{\sqcup}

\forcecommand{\phil}{\phi}

\forcecommand{\FVSE}{\mathsf{4VSE}}
\forcecommand{\Tern}{\mathsf{4VSE}}
\forcecommand{\BSE}{\mathrm{2VSE}}
\forcecommand{\tand}{ \: \& \: }
\forcecommand{\tlub}{ \: \lub \: }
\forcecommand{\tglb}{ \: \glb \: }
\forcecommand{\tstar}{ \: \star \:}

\forcecommand{\tor}{ \: \mathbf{+} \: } 

\forcecommand{\AndS}{\mathrm{\&_\mathrm{S}}}
\forcecommand{\OrS}{\mathrm{+_\mathrm{S}}}
\forcecommand{\LubS}{\mathrm{\lub_\mathrm{S}}}
\forcecommand{\WhenS}{\mathrm{\mapsto_\mathrm{S}}}

\forcecommand{\AndR}{\AndB}
\forcecommand{\OrR}{\OrB}
\forcecommand{\NotR}{\NotB}
\forcecommand{\AndB}{\mathrm{\bt{AND}}}
\forcecommand{\OrB}{\mathrm{\bt{OR}}}
\forcecommand{\NotB}{\mathrm{\bt{NOT}}}
\forcecommand{\AndF}{\mathrm{\ft{AND}}}
\forcecommand{\OrF}{\mathrm{\ft{OR}}}
\forcecommand{\NotF}{\mathrm{\ft{NOT}}}

\forcecommand{\la}{\langle}
\forcecommand{\ra}{\rangle}
\forcecommand{\dla}{\langle\langle}
\forcecommand{\dra}{\rangle\rangle}

\forcecommand{\lift}[1]{#1}

\forcecommand{\ds}[3]{\dsf{#1}(#2)(#3)}
\forcecommand{\dt}[3]{\dtf{#1}(#2)(#3)}

\forcecommand{\dsf}[1]{{}^{\phi}[\:#1\:]}
\forcecommand{\dtf}[1]{{}^{\phi}_F[\hspace{-0.4mm}[\;#1\;]\hspace{-0.4mm}]}

\forcecommand{\dsfc}[1]{[\:#1\:]}
\forcecommand{\dtfc}[1]{{}_{F}[\hspace{-0.4mm}[\;#1\;]\hspace{-0.4mm}]}

\forcecommand{\dstn}[1]{\ds{#1}{t}{n}}
\forcecommand{\dttn}[1]{\dt{#1}{t}{n}}
\forcecommand{\dse}[3]{\dsef{#1}(#2)(#3)}
\forcecommand{\dte}[3]{\dtef{#1}(#2)(#3)}
\forcecommand{\dsef}[1]{\langle{}#1\rangle{}}
\forcecommand{\dtef}[1]{{}_{\symf}\langle{}\langle{}#1\rangle{}\rangle{}}
\forcecommand{\dsetn}[1]{\dse{#1}{t}{n}}
\forcecommand{\dtetn}[1]{\dte{#1}{t}{n}}
\forcecommand{\wdse}[3]{{}^{\mathsf{DEF}}\dsef{#1}(#2)(#3)}
\forcecommand{\wdsetn}[1]{\wdse{#1}{t}{n}}

\forcecommand{\dnt}[3]{\langle{}#1\rangle(#3)(#2)}
\forcecommand{\dtn}[3]{\langle#1\rangle(#2)(#3)}

\forcecommand{\varA}[2]{a_{#2}^{#1}}
\forcecommand{\varC}[2]{c_{#2}^{#1}}
\forcecommand{\varD}[2]{d_{#2}^{#1}}
\forcecommand{\varK}[2]{k_{#2}^{#1}}
\forcecommand{\varAtn}{a_{t}^{n}}
\forcecommand{\varCtn}{c_{t}^{n}}
\forcecommand{\varDtn}{d_{t}^{n}}
\forcecommand{\varKtn}{k_{t}^{n}}

\forcecommand{\domainr}{\upharpoonright}

\forcecommand{\Nodes}{\mathcal{N}}
\forcecommand{\In}{\mathcal{I}}
\forcecommand{\Out}{\mathcal{O}}
\forcecommand{\State}{\mathcal{S}}

\forcecommand{\TEL}{\mathsf{TEL}}
\forcecommand{\GTEL}{\mathsf{GTEL}}
\forcecommand{\N}{\mathrm{\mathbf{N}}}
\forcecommand{\is}{\mathbf{\: is \:}}
\forcecommand{\en}{\mathbf{\: \: and \: \:}}
\forcecommand{\True}{{\cdot}}
\forcecommand{\Ass}{\mathrm{Ass}}
\forcecommand{\ass}{\Ass}

\forcecommand{\V}{\mathbb{V}}
\forcecommand{\D}{\mathbb{V}}
\forcecommand{\B}{\mathbb{B}}

\forcecommand{\seqp}{\mathbf{seq}}
\forcecommand{\Seq}{\mathbf{Seq}}
\forcecommand{\CircuitState}{\mathbf{State}}
\forcecommand{\circuitState}{\CircuitState}

\forcecommand{\Nat}{\mathbb{N}}
\forcecommand{\Lra}{\Longrightarrow}
 
\forcecommand{\Traj}{\mathsf{Traj}}
\forcecommand{\TrajF}[1]{\ft{\Traj}(#1)}
\forcecommand{\TrajB}[1]{\bt{\TrajB}(#1)}

\forcecommand{\lit}{\mathsf{lit}}
\forcecommand{\Var}{\mathsf{Vars}}

\forcecommand{\Imp}{\Longrightarrow}
\forcecommand{\lpt}{\preceq}

\forcecommand{\modTel}{\models}
\forcecommand{\modPropTwo}{\models_{\mathrm{Prop}}} 
\forcecommand{\modBool}{\models_{\mathrm{2}}}   
\forcecommand{\modBd}{\models_{\leftrightarrows}}  
\forcecommand{\modC}{\models_{\mathrm{Cautious}}}  
\forcecommand{\modS}{\models_{\mathrm{Simple}}}  
\forcecommand{\modFwdCautious}{\modFw^{\mathrm{cautious}}}
\forcecommand{\modFwd}{\models_{\rightarrow}}
\forcecommand{\modFw}{\modFwd}
\forcecommand{\modFwCautious}{\modFwdCautious}

\forcecommand{\modY}{\models_{\mathrm{Y}}}     

\forcecommand{\cnode}[1]{\mathsf{#1}}
\forcecommand{\nin}{\cnode{in}}
\forcecommand{\nset}{\cnode{set}}
\forcecommand{\np}{\cnode{p}}
\forcecommand{\npp}{\cnode{p'}}
\forcecommand{\nq}{\cnode{q}}
\forcecommand{\nr}{\cnode{r}}
\forcecommand{\ns}{\cnode{s}}
\forcecommand{\nuu}{\cnode{u}}
\forcecommand{\nv}{\cnode{v}}
\forcecommand{\nnset}{\neg\cnode{set}}
\forcecommand{\nout}{\cnode{out}}
\forcecommand{\noutp}{\cnode{out'}} 
\forcecommand{\nreg}{\cnode{reg}}
\forcecommand{\nregp}{\cnode{reg'}}

\forcecommand{\Lits}{\mathrm{Lits}}
\forcecommand{\spunion}{\cup}
\forcecommand{\union}{\cup}
\forcecommand{\co}{\{|}
\forcecommand{\cc}{|\}}

\forcecommand{\true}{\mathit{true}}

\forcecommand{\ce}{\leq}

%% file: abstract.tex
\begin{abstract} 
Generalised Symbolic Trajectory Evaluation (GSTE) is a high-capacity
formal verification technique for hardware. GSTE is an extension of
Symbolic Trajectory Evaluation (STE). The difference is that STE is
limited to properties ranging over finite time-intervals whereas GSTE
can deal with properties over unbounded time.

GSTE uses \emph{abstraction}, meaning that details of the circuit
behaviour are removed from the circuit model.  This improves the
capacity of the method, but has as down-side that certain properties
cannot be proven if the wrong abstraction is chosen.

A semantics for GSTE can be used to predict and understand why certain
circuit properties can or cannot be proven by GSTE. Several semantics
have been described for GSTE by Yang and Seger. 
These semantics, however, are not \emph{faithful} to the proving power
of GSTE-algorithms, that is, the GSTE-algorithms are \emph{incomplete}
with respect to the semantics. The reason is that these semantics do
not capture the abstraction used in GSTE precisely.


The abstraction used in GSTE makes it hard to understand why a
specific property can, or cannot, be proven by GSTE. The semantics
mentioned above cannot help the user in doing so.  So, in the current
situation, users of GSTE often have to revert to the GSTE algorithm to
understand why a property can or cannot be proven by GSTE.

The contribution of this paper is a \emph{faithful semantics} for
GSTE. That is, we give a simple formal theory that deems a property to
be true if-and-only-if the property can be proven by a GSTE-model
checker.  We prove that the GSTE algorithm is sound and complete with
respect to this semantics.  Furthermore, we show that our semantics
for GSTE is a generalisation of the semantics for STE and give a
number of additional properties relating the two semantics.

\end{abstract}

%% file: introduction.tex
\section{Introduction}


The rapid growth in hardware complexity has led to a need for \emph{formal verification}
of hardware designs to prevent bugs from entering the final silicon.
\emph{Model checking} is a verification method in which  a model of a system is checked
against a \emph{property},
describing the desired behaviour of the system over time. 
Today, all major hardware companies use model checkers in order to
reduce the number of bugs in their designs.

%
\subsection{Symbolic Trajectory Evaluation} 

\emph{Symbolic Trajectory Evaluation} (STE) \cite{ste} is a high-performance
model checking technique based on \emph{simulation}. STE combines
three-valued simulation (using the standard values 0 and 1 together with the
extra value $\X$, ``don't know") with symbolic simulation (using symbolic
expressions to drive inputs). STE has been extremely successful in verifying
properties of circuits containing large data paths (such as memories, FIFOs,
and floating point units) that are beyond the reach of traditional symbolic
model checking \cite{methodology,highlevel,ste}.

\begin{figure}[t]
\[
\epsfig{file=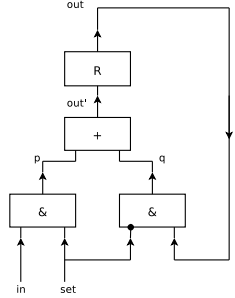,width = 3cm}
\]
\caption{A memory cell}
\label{fig:memorycellGSTE}
\end{figure}

Consider the circuit in Figure \ref{fig:memorycellGSTE}.
The circuit
consists of two AND-gates, an OR-gate, a register (depicted by the letter R), and an inverter (depicted by the black dot). 
The register has output node
$\nreg$ and input node $\nregp$. The value of the output of the register at time
$t+1$ is the value of its input at time $t$. 
The memory cell can be written with the value at node $\nin$ by making node $\nset$ high.

In STE, circuit specifications are \emph{assertions} of the form $A \Longrightarrow C$. Here, $A$ is called the
\emph{antecedent} and $C$ the \emph{consequent}. 
For example, an STE-assertion for the memory cell is:
\[
(\nin \is a) \en (\nset \is 1) \Longrightarrow \N(\nout \is a)
\]
Here $a$ is a \emph{symbolic constant}\footnote{
The name \emph{symbolic constant} is used to indicate that the variable keeps
a constant value over different points in time. In plain STE, such variables
are called \emph{symbolic variables}. As this paper deals with GSTE, we will
use the GSTE terminology even when we discuss plain STE.
}
, which can take on the value $0$ or $1$, and $\nin$,
$\nset$ and $\nout$ are \emph{node names}. $\N$ is the next-time operator. The assertion states
that when node $\nin$ has value $a$, and node $\nset$ has value $1$, then at the next point in
time, node $\nout$ must have value $a$.

\subsection{Generalised Symbolic Trajectory Evaluation}
One of the main disadvantages of STE is that it can only deal 
with properties ranging over a finite number of time-steps.
\emph{Generalised Symbolic Trajectory Evaluation} (GSTE) 
\cite{gsteIntroduction,gsteCaseStudy,abstractionInAction,gsteNotPublished} 
is an extension of STE that can deal with properties ranging over unbounded time.

In GSTE, circuit properties are given by \emph{assertion graphs}.
For example, an assertion graph for the memory cell is:
\begin{equation}
\xymatrix{ 
{\init}  \ar[rrrr]_{(\nin \is a) \en (\nset \is 1) / \True} &&&& v \ar@(ul,ur)^{\nset \is 0/ \True} 
\ar[rrrr]_{\True / \nout \is a}
&&&& w \ar@(ur,dr) 
}
\label{ag_mc1}
\end{equation}
In the assertion graph, each edge is labelled with a pair $A/C$. As in STE,
$A$ is called the antecedent and $C$ is called the consequent.  
The syntax of $A$ and $C$ is like the syntax of the antecedent
and consequent in STE without the next-time operator $\N$. The
$\N$ operator can not be used because each edge only represents a single time-point.
A dot $(\cdot)$ means an empty antecedent or consequent.

The assertion graph above states that if 
we write
value $a$ to the memory cell, and then for arbitrary many time-steps
we do not write, the memory cell still contains value $a$.

Each finite path, starting in the initial vertex $\init$ of the graph,
represents an STE property. For instance, the finite paths through the assertion graph above
represent the following STE properties:
\[
\begin{array}{lcl}
(\nin \is a) \en (\nset \is 1) & \Longrightarrow & \N(\nout \is a) \\
(\nin \is a) \en (\nset \is 1) \en \N(\nset \is 0) & \Longrightarrow & \N\N(\nout \is a) \\
(\nin \is a) \en (\nset \is 1) \en \N(\nset \is 0) \en \N\N(\nset \is 0) & \Longrightarrow & \N\N\N(\nout \is a)  \\
\ldots
\end{array}
\] 
Each of these assertions can be proven by an STE model checker. But, as the set of assertions
is infinite, we cannot use plain STE to prove all of them.
%
%
However, if  we use GSTE to prove that the circuit satisfies the above assertion graph, it follows
that all STE-assertions represented by the assertion graph hold as well.

Note that in GSTE, just like in STE, the initial values of registers are ignored.


\subsection{Earlier work on semantics for GSTE}
A \emph{semantics} for GSTE can be used to predict and understand why certain circuit properties
can or cannot be proven by GSTE. In \cite{gsteNotPublished,abstractionInAction} three semantics for GSTE are distinguished: 
(1) the \emph{strong semantics},
(2) the \emph{normal semantics}, and 
(3) the \emph{fair semantics}.
The semantics have in common that a circuits satisfies an assertion-graph if
it satisfies all \emph{appropriate} paths in the assertion graph. The meaning
of appropriate differs over the three semantics, as we explain in the
following paragraphs. As in \cite{reasoning}, we refer to this class of
semantics as the \emph{$\forall$-semantics}, because these semantics really
consider {\em all} concrete paths, rather than approximating this
quantification by applying abstraction.

In the strong semantics, a circuit satisfies a GSTE assertion graph if-and-only-if 
the circuit satisfies all STE-assertions corresponding to \emph{finite} paths
in the assertion graph. 
For instance, as the memory cell
satisfies the set of finite assertions above, it also satisfies assertion graph (\ref{ag_mc1}).

Consider the following assertion graph:
\begin{equation}
\xymatrix{ 
{\init}  \ar[rrrr]_{(\nset \is 1) / (\nin \is a)  } &&&& v 
\ar[rrrr]_{\nout \is a / \True}
&&&& w \ar@(ur,dr) 
}
\label{ag_mc2}
\end{equation}
Intuitively, we might want the above assertion graph to state
that if at some time-point node $\nout$ has
value $a$, and just before that, node $\nset$ was high, then at 
this time-point node $\nin$ should have value $a$. This is an example 
of a \emph{backwards property}, that is, a property in which 
a consequent depends on an antecedent at a later time-point.

The strong semantics cannot deal with such backwards properties. 
For instance, for the above property, the path starting
in vertex $\init$ and ending in vertex $v$ corresponds to the assertion
\[
(\nset \is 1) \Longrightarrow (\nin \is a)  
\]
This assertion is, of course, not true for the memory cell. 
But, any run of the circuit that makes $\nin \is a$ fail, 
makes $\N(\nout \is a)$ fail as well.
So, intuitively, the assertion is not satisfied 
because a consequent failed before the antecedent it depended on
could fail. 

In the normal semantics, a circuit satisfies a GSTE assertion graph if-and-only-if 
the circuit satisfies the STE-assertions corresponding to all \emph{infinite} paths 
in the assertion graph.
Therefore, the normal semantics can deal with backwards properties as well.
For instance, in assertion graph (\ref{ag_mc2}), there is only one infinite path.
This path 
corresponds to the following assertion:
\[
(\nset \is 1) \en \N(\nout \is a) \Longrightarrow (\nin \is a)  
\]
As any circuit trace that satisfies the antecedent satisfies the consequent as well,
this assertion is satisfied
by the circuit. Thus, in the normal semantics, the GSTE assertion graph is satisfied.

Finally, the need for the fair semantics is illustrated by the following example.
Consider the assertion graph:
\begin{equation}
\xymatrix{ 
{\init}  \ar[rrrr]_{(\nset \is 1) / (\nin \is a)  } &&&& v 
\ar@(ul,ur)^{\nset \is 0/ \True} 
\ar[rrrr]_{\nout \is a / \True}
&&&& w \ar@(ur,dr) 
}
\label{ag_mc3}
\end{equation}
The assertion graph above states that if at some time-point node $\nout$ has
value $a$, and before that, for a period of time no values were written
to the memory-cell, and before that, $\nset$ was high, then at 
this time-point $\nin$ should have value $a$.  

In the normal semantics, the memory cell circuit does not satisfy
this assertion graph. Consider the infinite path starting in $\init$
and then cycling at the self-loop at $v$ for ever. This path corresponds
to the infinite assertion:
\[
(\nset \is 1) \en \N(\nset \is 0) \en \N\N(\nset \is 0) 
 \en \ldots \Longrightarrow \nin \is a  
\]
For a given $a$, this assertion can be falsified by the trace in which value 
$\neg a$ is written at time 0, and is kept in memory since then.

In the fair semantics for GSTE, this problem is solved by selecting a set
of \emph{fair edges}. The semantics only considers paths that
visit every fair edge infinitely often.
For instance, if in the above assertion graph
the edge from vertex $w$ to itself is made fair, the assertion graph holds in the fair semantics.

\subsection{GSTE model checking} 
In the same papers~\cite{gsteNotPublished,abstractionInAction}, model checking algorithms for normal, strong and
fair GSTE are described.
%
%
%
%
It is proven that the model checking algorithms are sound with respect to their corresponding
semantics. However, the algorithms are not \emph{complete}. The reason is that the $\forall$-semantics 
do not precisely capture the information loss due to the three-valued abstraction in GSTE.



%
For example, consider the following circuit
\[
\epsfig{file=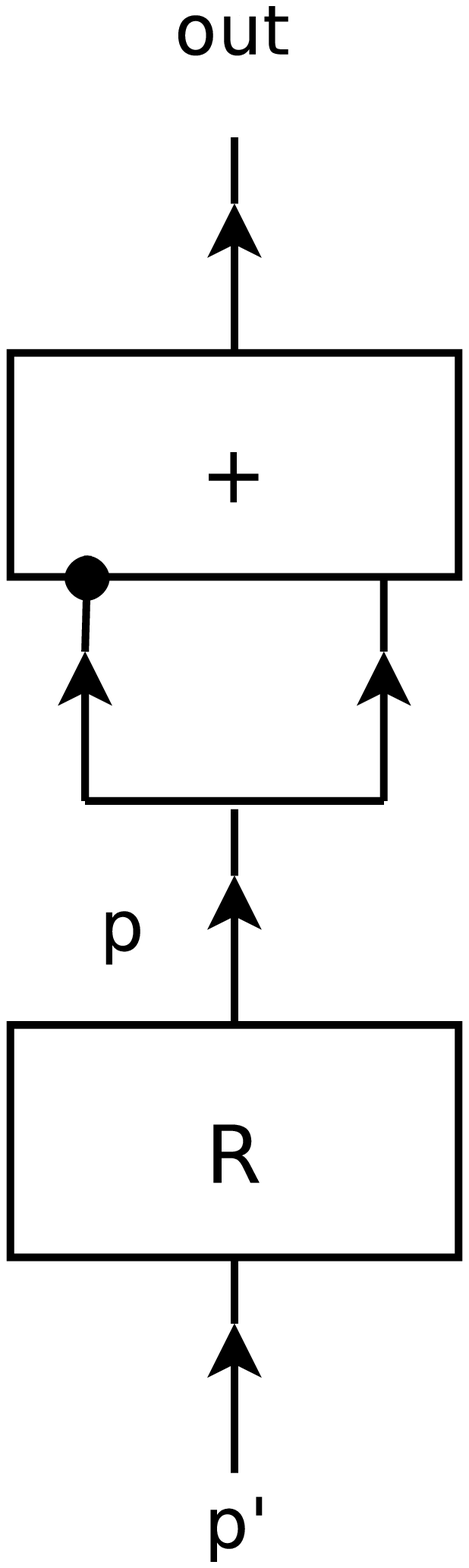,width = 2cm}
\]
and the following assertion graph
\[
\xymatrix{ 
{\init} \ar@/^/[rr]^{\npp \is 1 / \True} \ar@/_/[rr]_{\npp \is 0 / \True}   
&& v \ar[rr]^{\True / \nout \is 1}   
&& w \ar@(ur,dr) 
}
\]
The assertion graph represents the following STE-assertions:
\[
\begin{array}{lcl}
\npp \is 1 & \Longrightarrow & \N (\nout \is 1) \\
\npp \is 0 & \Longrightarrow & \N (\nout \is 1) \\
\end{array}
\] 
Both assertions hold. So, the semantics described above   predict that the circuit satisfies the assertion graph.

However, it turns out that the GSTE-algorithm cannot prove the assertion
graph! The reason is that GSTE algorithms only compute one three-valued
assertion for each edge in the assertion graph. This is in general not enough
to take account for all STE assertions corresponding to all paths through the
assertion graph, so a certain {\em information loss} happens. In this
particular case, the state calculated on the edge from $v$ to $w$ gives value
$\X$ to node $\nout$. This can be explained as follows. The antecedent at the
top edge between vertices $\init$ and $v$ requires node $\npp$ to have value
$1$. The antecedent at the bottom edge requires node $\npp$ to have value
$0$. Node $\npp$ is the input to a register with node $\np$ as output. So,
when the edge from $v$ to $w$ is reached via the top edge between $\init$ and
$v$ node $\np$ will receive value $1$. When the edge from $v$ to $w$ is
reached via the bottom edge between $\init$ and $v$, node $\np$ will receive
value $0$. As the value of node $\np$ should comply with both paths, the
algorithm chooses value $\X$ for node $\np$, and thus node $\nout$ receives
value $\X$ as well.



\subsection{The problem}
The previous example illustrates that the
$\forall$-semantics for GSTE discussed previously cannot be used to explain
how the three-valued abstraction causes certain properties to be not provable
with GSTE. This can lead to situations where seemingly trivial changes to either
the circuit or the assertion can suddenly make an assertion not provable anymore.


This is an undesirable situation. We believe that a faithful semantics for GSTE is
needed. 

A faithful semantics deems a property to be true if-and-only-if the property
can be proven by a GSTE-model checker. Without a faithful semantics, a GSTE
verification engineer is left to the particular internals of the model
checker at hand to understand what can and cannot be proved. Also, a faithful
semantics can be used to understand differences between different GSTE model
checkers. For example, the GSTE semantics of satGSTE \cite{GSTESAT} is
expressed using successive unrollings of the assertion graph as STE
assertions. However, the abstraction obtained in that way does not correspond
to the abstraction in standard GSTE model checkers. This means that there are
assertion graphs for which satGSTE and standard GSTE model checkers give
different answers.

To further clarify the importance of a faithful GSTE semantics, we would like
to point out that there is a difference between the use of abstraction in
(G)STE, and the application of abstraction as a performance enhancer in model
checkers for standard temporal logics like LTL and CTL. In the latter case, a
model checker might simply give up when it happens to choose an abstraction
that is too weak to prove a property, but it is still clear to the
verification engineer what the specification means. In (G)STE, what
abstraction to use in the model checker is an {\em artefact of the
specification}, not an artefact of the model checker. So, in (G)STE it is
vital to understand what a specification means, separate from a particular
model checker, including the abstraction that is specified.

In previous work~\cite{steFaithful}, we have described a faithful semantics for STE.
However, up till now, no faithful semantics for GSTE has been described.

\subsection{Our contribution}
In this paper, we present a semantics for GSTE that is faithful to the proving
power of the GSTE model checking algorithm. 
Compared to the semantics described in \cite{gsteNotPublished,abstractionInAction}, our semantics 
corresponds to the strong semantics of GSTE. That is, in this paper, we do not consider
backwards properties or fairness constraints, which remains future work. 
One difference with the strong semantics in
\cite{gsteNotPublished,abstractionInAction} is that our
semantics captures the three-valued abstraction of GSTE precisely, and thus
can be used to explain the information loss caused by the three-valued
abstraction in GSTE. 

Another difference is that our semantics for GSTE follows the same structure
as the semantics for STE~\cite{steFaithful,ste,indexing}. For instance, where STE deals with \emph{sequences}
to represent abstract circuit behaviour, our GSTE semantics uses
\emph{sequence graphs}. Here, a sequence graph is a mapping from edges 
in an assertion graph to abstract circuit states.
%
%
We show that our GSTE semantics is
a generalisation of the STE semantics. That is, given a linear
assertion graph, the STE-semantics and GSTE-semantics are equivalent.
Finally, we state a number of additional properties relating the two semantics.

We believe that our faithful semantics for STE is an important contribution to 
the research on GSTE for at least two reasons.

First of all, a faithful semantics makes GSTE more accessible to novice users: 
a faithful semantics enables users to understand the abstraction used in GSTE,
without having to understand the details of the model checking algorithm.
Additionally, in
this paper, we aim at increasing the understanding for GSTE users of subtle cases of
information-loss due to abstraction by providing enlightening examples.

Furthermore, 
a faithful semantics for GSTE can be used
as basis for research on new GSTE model checking algorithms and other GSTE tools.
To illustrate this, in previous work~\cite{roorda}, we described a new SAT-based model checking algorithm
for STE and proven that it is sound and complete with respect to our faithful semantics
for STE presented in~\cite{steFaithful}. Without a faithful semantics for STE, we would have been forced to prove the
correctness of our algorithm by relating it to other model checking algorithms
for STE. This is clearly a more involved and less elegant approach. 
In fact, we believe that without constructing a faithful semantics for STE first,
we would not have obtained the level of understanding of STE needed to develop the
new SAT-based model checking algorithm. 

In the same way, we expect that the faithful semantics for GSTE presented
in this paper will open the door for new research on GSTE model checking algorithms and other GSTE tools.

\subsection{Other related work}

The following papers are based on the $\forall$-semantics
for GSTE.

\subsubsection*{GSTE as partitioned model checking}

In \cite{vardi}, the relation between GSTE and 
classic symbolic model checking is studied.
It is explained how GSTE can be seen
as a \emph{partitioned} form of classic symbolic model checking.
However, the abstraction of GSTE is not
taken into account. Therefore, this paper,
focussing on the abstraction in GSTE,
is complementary to \cite{vardi}. 

\subsubsection*{Using SAT for debugging of GSTE assertion graphs}
In \cite{GSTESAT}, the tool \emph{satGSTE} is presented.
The tool considers a finite subset of all   finite paths in an
assertion graphs, for instance, all paths up to a certain length.
For each path in this subset, the tool model checks the corresponding STE assertion.
The authors explain how the tool can be used to debug 
and refine GSTE assertion graphs. However, their tool does not follow the same
semantics as standard GSTE model checking algorithms. Thus, certain counter
examples that would occur in a standard GSTE model checker due to the use of
abstraction cannot be found with their algorithm.

\subsubsection*{Monitor circuits for GSTE assertion graphs}

In (conventional, non-symbolic) simulation, a model
of a circuit is fed with a large number of inputs. For
every input it is checked whether the output is as expected.
Typically, a \emph{monitor circuit} is used to make this check.
The monitor circuit observes the system under verification without interfering. 
During each step of the simulation, it indicates whether the system has obeyed 
the formal specification  thus far.

In \cite{monitor1,monitor2} methods for automatic
construction of monitor circuits for GSTE assertion graphs are described.
The method in \cite{monitor1} requires the use of a symbolic simulator
if the assertion graph contains symbolic constants.
In \cite{monitor2} it is explained how,
for the class of so-called \emph{simulation friendly assertion graphs},
the method of \cite{monitor1} can be extended to deal with symbolic constants
even in conventional non-symbolic simulation.

The papers explain how monitor circuits
can be used to make a bridge between GSTE model checking and
conventional simulation. For instance, monitor circuits
can be used  to quickly debug and refine GSTE specifications before trying to use
more labour intensive GSTE model checking. 

%

\subsubsection*{Reasoning about GSTE assertion graphs}

Using the construction of monitor circuits for GSTE assertion
graphs, \cite{reasoning} describes two algorithms that can be
used in compositional verification using GSTE. 
The first algorithm decides whether one assertion graph
implies another.
The second algorithm can be used to model check an assertion
graph under the assumption that another assertion graph is true.


\subsubsection{Relation to this paper} 
Each of the papers above is based on the $\forall$-semantics for GSTE.
As explained above, the $\forall$-semantics are not faithful to
the proving power of the GSTE model checking algorithms.
So, it can occur that a tool described in the papers deems a 
GSTE assertion to be true, while the GSTE model checking
algorithm cannot prove it.

For instance, the monitor circuits described above cannot
be used to debug and refine assertions graphs that are true in the
$\forall$-semantics but yield a spurious counter-example
 when trying to prove them with a GSTE model checker.
The satGSTE tool is limited in the same way. 
We elaborate further on this in the future work section of this paper.

\subsection{Structure of this paper}
In the next section, we revisit the semantics of STE assertions.
Then, in Section 3, we present our semantics of GSTE assertion graphs.
In Section 4, we compare the STE semantics with the 
GSTE semantics by giving a number of properties describing
their relation. In Section 5, we describe the GSTE model checking
algorithm and show that it is sound and complete with respect
to our semantics. Finally, in Section 6, we conclude and give
suggestions for future work.

%% file: ste.tex
\section{STE Preliminaries}

A semantics for STE was first described by Seger and Bryant~\cite{ste}.
Later, a simplified and easier to understand semantics was given
by Melham and Jones~\cite{indexing}. Both of these semantics are expressed in
terms of a {\em next state function}, expressing the relationship between two
consecutive states in the circuit. Unfortunately, neither of these semantics
matches the proving power of currently available STE model checkers.
The problem is that they cannot deal with \emph{combinational} properties
(properties ranging over one single point in time). All such properties are
deemed to be false by the semantics. Therefore standard next state semantics
does not seem to be a good starting point for finding a faithful semantics for
GSTE.

In previous work~\cite{steFaithful}, we have described an alternative
semantics for STE that actually is faithful to the proving power of STE
model checkers. The semantics is called the \emph{closure semantics}.
Informally, the closure semantics only differs from the traditional STE
semantics for combinational properties.

A main ingredient of the closure semantics for STE is the concept of a
\emph{closure function}. The idea is that a closure function takes as input a
state of the circuit, and calculates all information about the circuit state
at the same point in time that can be derived by propagating the information
in the input state in a forwards fashion.
In the next section, we give an alternative semantics for GSTE also based on
closure functions.

In this section we briefly describe the closure semantics for STE. For more
examples and a discussion on the differences with the semantics given in
\cite{ste,indexing}, we refer the reader to \cite{steFaithful}.

Readers familiar with \cite{steFaithful} can skip most of this section;
compared to \cite{steFaithful} we slightly changed notation in the definition
of the closure function on sequences, and we introduced an extra variant of a
closure semantics called the simple semantics. Furthermore, we adapted the
terminology to GSTE: we call the variables in STE-assertions \emph{symbolic
constants} to indicate that they keep a constant value over time.
Finally, we use \emph{finite} sequences to represent
circuit behaviour, as opposed to the standard use of infinite sequences.
Notice that this is a very superficial change on the notational level; it
does not change the semantics itself.
The reason for making the change is that it enables us to considerably
simplify the proof of Proposition \ref{prop:compare} on page
\pageref{prop:compare}.

\forcecommand{\edge}[1]{\ar@{-}[#1]}
\forcecommand{\node}{*+[o][F-]{ }}

\begin{figure}[t]
\centerline{
\xymatrix{
                         & \T \\
0 \edge{ur}\edge{dr} &   & 1 \edge{ul} \edge{dl}   \\
                         & \X
}
}
\caption{The STE lattice}
\label{fig:lattice}
\end{figure}
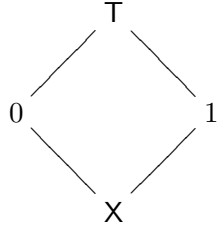

\subsection{Values and Circuits States}

\head{Values} In STE, we can abstract away from specific Boolean values of a
node, by using the value $\X$, which stands for
\emph{unknown}. The value $\T$ stands for \emph{over constrained}. A node takes
on the value $\T$ when is required to have both value $0$ and value $1$. 
%

On this set an \emph{information-ordering} $\leq$ is introduced, 
see Figure \ref{fig:lattice}. 
The unknown value $\X$ contains the least information, so  $\X \leq 0$ and $\X \leq 1$,
while $0$ and $1$ are incomparable.
The over-constrained value contains the most
information, so $0 \leq \T$ and $1 \leq \T$.  
If $v \leq w$ it is said that $v$ is \emph{weaker} 
than $w$.   

\setlength\arraycolsep{1.4pt}
\setlength\tabcolsep{1.4pt}

\begin{figure}[t]
\[
\begin{array}{ccccccccc}
\begin{array}{c|c}
v    & \neg v \\
\hline
0    & 1    \\
1    & 0    \\
\X   & \X    \\
\T   & \T   \\
\end{array}
& 
  \hspace{0.7cm}
&
\begin{array}{c|cccc}
\tand& 0  & 1  & \X & \T \\
\hline
0    & 0  & 0  & 0  & \T \\
1    & 0  & 1  & \X & \T \\
\X   & 0  & \X & \X & \T \\
\T   & \T & \T & \T & \T \\
\end{array}
& 
  \hspace{0.7cm}
&
\begin{array}{c|cccc}
\tor & 0  & 1  & \X & \T \\
\hline
0    & 0  & 1  & \X  & \T \\
1    & 1  & 1  & 1  & \T \\
\X   & \X & 1  & \X & \T \\
\T   & \T & \T & \T & \T \\
\end{array}
& 
  \hspace{0.7cm}
&
\begin{array}{c|cccc}
\lub & 0  & 1  & \X & \T \\
\hline
0    & 0  & \T & 0  & \T \\
1    & \T & 1  & 1  & \T \\
\X   & 0  & 1  & \X & \T \\
\T   & \T & \T & \T & \T \\
\end{array}
& 
  \hspace{0.7cm}
&
\begin{array}{c|cccc}
\glb & 0  & 1  & \X & \T \\
\hline
0    & 0  & \X & \X & 0 \\
1    & \X & 1  & \X & 1 \\
\X   & \X & \X & \X & \X \\
\T   & 0  & 1  & \X & \T \\
\end{array}
\end{array}
\]
\caption{Four-valued extensions of the logical operators, least upper bound and greatest lower bound operators.} 
\label{fig:extensionsGS}
\end{figure}

The set $V$ together with the ordering $\leq$ forms a \emph{lattice}.
The \emph{least upper bound operator} is written $\lub$,
the \emph{greatest lower bound operator} is written $\glb$, see Figure
\ref{fig:extensionsGS}.

The logical operators for conjunction, written $\tand$,
disjunction, written $\tor$, and negation, written $\neg$, are
extended to the four-valued domain as in Figure \ref{fig:extensionsGS}.

\head{States} A \emph{circuit state}, written $s : \CircuitState$, is a
function from the set of nodes of a circuit to the values
$\{0,1,\X,\T\}$\footnote{Such an STE circuit state can be thought of as
representing a {\em set} of regular states, commonly used in set-based
abstractions, where $\X$ represents the set $\{0,1\}$ and $\T$ represents the
empty set. This view induces a natural set-theoretic lattice, with set inclusion as
its ordering. It is perhaps confusing that the standard STE lattice ordering
(also used here) goes exactly the other way around; i.e. the STE $\lub$
corresponds to $\cap$ and $\glb$ corresponds to $\cup$.}.

\subsection{Closure functions}
\label{sec:cfs}
In our semantics for STE, \emph{closure functions} are used as 
circuit models. The idea is that a closure function, written $F : \CircuitState \rightarrow \CircuitState$ 
takes as input a state of the circuit,  and calculates all information about the circuit state
at the \emph{same} point in time that can be derived by propagating the information in the
input state in a \emph{forwards} fashion.

\begin{exa}
The closure function for a circuit consisting of a single AND-gate with inputs
$\np$ and $\nq$, and output $\nr$ is given by the table below. Here, $s$ is a state and
$n$ is a node.
\[
\begin{array}{l|l}
n    & F(s)(n) \\
\hline
\np    & s(\np)   \\ 
\nq    & s(\nq)   \\
\nr    & (s(\np) \tand s(\nq)) \lub s(\nr) \\
\end{array} 
\]
The least upper bound operator in the expression for $F(s)(\nr)$ combines
the value of $\nr$ in the given state $s$, and the value for $\nr$ that
can be derived from the values of $\np$ and $\nq$, being $s(\np) \tand s(\nq)$. 

A state $s : \{\np,\nq,\nr\} \rightarrow \V$ can be written as a vector
$s(\np),s(\nq),s(\nr)$. For example, the state that assigns the value 1 to nodes $\np$ and $\nq$
and the value $\X$ to node $\nr$ is written as $11\X$.
Applying the closure function to the state $11\X$ yields $111$. The reason is that when
both inputs to the AND-gate have value $1$, then by forwards propagation of information, also the output has value $1$. 
Applying the
closure function to state $1\X\X$ yields $1\X\X$. The reason  is that the output of
the AND-gate is unknown when one input has value $1$ and the other value $\X$. The
\emph{forwards} nature of simulation becomes clear when the closure function is applied
to state $\X\X1$, resulting in $\X\X1$. Although the inputs to the
AND-gate must have value $1$ when the output of the gate has value $1$, this cannot be derived by
forwards propagation.  

A final example shows how the over-constrained value $\T$ can arise. Applying the
closure function to state $0\X1$ yields $0\X\T$. 
The reason  is that the input state gives node $\nr$ value $1$ and node $\np$
value $0$.
From $\np$ having value $0$ it can be derived by forwards propagation that  $\nr$ has value $0$,
therefore $\nr$ receives the over-constrained value $\T$. \qed
\end{exa}
A closure function is a function $F : \CircuitState \rightarrow
\CircuitState$ satisfying the following three conditions:
\begin{enumerate}[$\bullet$]
\item
$F$ is \emph{monotonic},  that is, for all
states $s_1,s_2$: $s_1 \leq s_2$ implies $F(s_1) \leq F(s_2)$. This means that a more
specified input state cannot lead to a less specified result. The reason is that given
a more specified input state, more information about the state of the circuit can be
derived.

\item
$F$ is \emph{idempotent}, that is,
for every state $s$: $F(F(s)) = F(s)$. This means that
repeated application of the closure function has the same result as applying the
function once. The reason is that the closure function should derive all
information about the circuit state in one go.
\item
$F$ is \emph{extensive},
that is, for every state $s$: $s \leq F(s)$. This means that
the application of a closure function to a circuit state should yield a state at least
as specified as the input state. The reason is that the closure function is required
not to lose any information.
\end{enumerate}

%
\head{Netlists}
Here, a
netlist is an acyclic list of definitions describing the relations between the
values of the nodes. 
Inverters are not modelled
explicitly in our netlists, instead they occur implicitly for each mention of
the negation operator $\neg{}$ on the inputs of the gates.
Registers are not mentioned explicitly in the netlist either. Instead,
for a register with output node $n$ in the circuit, the input of the register is
node $n'$ which is mentioned in the netlist. 
%
%
For simplicity, 
we only allow AND-gates and OR-gates in netlists.
It is, however, straightforward to extend this notion of netlists to include
more operations.

\head{Induced Closure Function} Given the netlist of a circuit $c$, the
\emph{induced closure function} for the circuit, written $F_c$, can easily be
constructed by interpreting each definition in the netlist as a four-valued
gate (see Figure \ref{fig:extensionsGS}). Each 

Given a state $s$, a circuit $c$, and a circuit node $n$, we calculate $F_c(s)(n)$ as follows:
\begin{enumerate}[$\bullet$]
\item If $n$ is a circuit input or the output of a register, then we define $F_c(s)(n) = s(n)$.
\item If $n$ is the output of an AND-gate with input nodes $p$ and $q$, then we define
$$
F_c(s)(n) = (F_c(s)(p) \tand F_c(s)(q)) \lub s(n).
$$
\item If $n$ is the output of an OR-gate with input nodes $p$ and $q$, then we define
$$
F_c(s)(n) = (F_c(s)(p) \tor F_c(s)(q)) \lub s(n).
$$
\item If $n$ is the output of an inverter with input node $p$, then we define
$$
F_c(s)(n) = \neg F_c(s)(p) \lub s(n).
$$
\end{enumerate}
This
definition is well-defined because netlists are acyclic by definition.

%
%

%

\begin{prop}
The induced closure function for a circuit is by construction monotonic, idempotent and extensive.
\end{prop}

\proof
The closure function $F_c$ is a composition of the monotonic functions
of four-valued negation,  four-valued conjunction and least upper bound, therefore it is
monotonic itself.

As netlists are acyclic by definition, we can prove properties by induction over the
definition of a node. We prove idempotency by proving $F_c(F_c(s))(n) = F_c(s)(n)$ by
induction on the definition of $n$. Assume $n$ is  in the set of input- and state-holding
nodes $\In \cup \State$, then $F_c(F_c(s))(n)=F_c(s)(n)$ by definition. If $n$ is defined
by $n = p \andg q$,   then: 
\[
\begin{array}{lll}
  &  F_c(F_c(s))(n)                                                                    \\
= & (F_c(F_c(s))(p) \tand F_c(F_c(s))(q)) \tlub F_c(s)(n)   & ($definition$)            \\
= & (F_c(s)(p) \tand F_c(s)(q)) \tlub F_c(s)(n)             & ($ind. hyp.$)  \\
= & (F_c(s)(p) \tand F_c(s)(q)) \tlub (F_c(s)(p) \tand F_c(s)(q)) \tlub s(n) & ($definition$)            \\
= & (F_c(s)(p) \tand F_c(s)(q)) \tlub s(n) &                                  ($property $ \lub)            \\
= &  F_c(s)(n) & ($definition$) 
\end{array} 
\]
A similar argument holds when $n$ is defined by a different gate definition.

The extensivity of $F_c$ follows directly from its definition: If $n$ is an input or state
holding node then $F_c(s)(n)=s(n)$, otherwise $F(c)(n)$ is defined as the least upper
bound of $s(n)$ and another expression, so $s(n) \leq F_c(s)(n)$. \qed

\subsection{A closure function for sequences}

\head{Sequences} A {\em sequence of depth $d$}, written
$\sigma : \{0,1,\ldots,d\} \rightarrow \CircuitState$, is a function from a point
in time to a circuit state, describing the behaviour of a circuit over time. The
set of all sequences is written $\Seq$. A \emph{three-valued sequence}
is a sequence that does not assign the value $\T$ to any node at any time.

The order $\leq$ and the operators $\lub$ and $\glb$ are extended 
to sequences in a point-wise fashion. That is, the order $\leq$ on sequences is defined by
$\sigma_1 \leq \sigma_2$ iff for all
$n$, $\sigma_1(n) \leq \sigma_2(n)$. Furthermore, $(\sigma_1 \lub \sigma_2)(n)=(\sigma_1(n) \lub \sigma_2(n))$,
and $(\sigma_1 \glb \sigma_2)(n)=(\sigma_1(n) \glb \sigma_2(n))$.

\head{Closure for sequences}
In STE, a circuit is simulated over multiple time steps. During simulation, information is
propagated forwards through the circuit and  through time, from each time step $t$ to time
step $t+1$. Note that the initial values of registers are ignored.

To model this forwards propagation of information through time, a  \emph{closure
function for sequences}, notation $\ftF : \Seq \rightarrow \Seq$, is used. Given a sequence,
the closure function for sequences calculates all information that can be derived
from that sequence by forwards propagation.
The closure function for sequences preserves the depth of the given sequence.

Recall that for every register with output $n$, the input to the register is node $n'$.
Therefore, the value of node $n'$ at time $t$ is propagated to node $n$ at time $t+1$
in the forwards closure for sequences.

Given a circuit state $s$, the function $\nextf$ calculates the information that
is propagated by the registers, and is defined by:
\[
\nextf(s)(n)  =  \twocases{s(n')}{n \in \State}{\X}{\text{otherwise}}      
\] 
%
The closure function for sequences $\Fseq$ is defined in terms of a closure
function $F$. 
Given a closure function $F$ for a circuit 
with a set of outputs of registers $\State$,
the \emph{closure function for sequences}, written $\ftF : \Seq \rightarrow
\Seq$, is inductively defined by:  
\[
\begin{array}{lcllr}
\ftF(\sigma)(0)       & = & F(\sigma(0))  &       \\
\ftF(\sigma)(t+1)     & = & F( \: \sigma(t+1) \tlub \nextf(\ftF(\sigma)(t)) \: )  & \hspace{0.5cm} & (0 \leq t \leq d-1)    \\
\end{array}
\]
\label{def:Fseq}
 %
%
%

\begin{prop}
The function $\ftF$ inherits the properties of being monotonic, idempotent and extensive from $F$. 
\end{prop}

\proof
The closure function $\ftF$ is a composition of the monotonic functions, $F$ and least
upper bound, therefore it is monotonic itself.

We prove the idempotency of $\ftF$ by proving $\ftF(\ftF(\sigma))(t)=\ftF(\sigma)(t)$ by
induction on $t$.

Suppose $t=0$, then
\[
\begin{array}{lll}
  & \ftF(\ftF(\sigma))(0)   \\
= & F(\ftF(\sigma)(0))     & ($definition of $ \ftF)  \\
= & F(F(\sigma(0))        & ($definition of $ \ftF)) \\
= & F(\sigma(0))          & ($idempotency of $F)    \\
= & \ftF(\sigma)(0)      & ($definition of $ \ftF)) \\
\end{array}
\]
The induction hypothesis is: $\ftF(\ftF(\sigma))(t)=\ftF(\sigma)(t)$
 for a fixed $t$. Suppose that the induction hypothesis holds, then:
\[
\begin{array}{lll}
  & \ftF(\ftF(\sigma))(t+1)     \\
= & F(\ftF(\sigma)(t+1) \tlub \nextf(\ftF(\ftF(\sigma))(t)) \:) & ($definition of $ \ftF) \\
= & F(\ftF(\sigma)(t+1) \tlub \nextf(\ftF(\sigma)(t)) \:)       & \text{(ind. hyp.)} \\
\end{array} 
\]
Now we reduce the term $\ftF(\sigma)(t+1) \tlub \nextf(\ftF(\sigma)(t))$ further.
\[
\begin{array}{lll}
  & \ftF(\sigma)(t+1) \tlub \nextf(\ftF(\sigma)(t))   &                       \\
= & F( \: \sigma(t+1) \tlub \nextf(\ftF(\sigma)(t))) \tlub  \nextf(\ftF(\sigma)(t)) & \text{(def. $\ftF$)} \\
= & F( \: \sigma(t+1) \tlub \nextf(\ftF(\sigma)(t))) & \text{($F$ extensive, prop. $\lub$)}
\end{array}
\]
Thus:
\[
\begin{array}{lll}
  & \ftF(\ftF(\sigma))(t+1)     \\
= & F(\ftF(\sigma)(t+1) \tlub \nextf(\ftF(\sigma)(t)) \:)       & \text{(see above)} \\
= & F(F( \: \sigma(t+1) \tlub \nextf(\ftF(\sigma)(t))))         & \text{(see above)} \\
= & F( \: \sigma(t+1) \tlub \nextf(\ftF(\sigma)(t)))            & \text{$F$ idempotent} \\
= & \ftF(\sigma)(t+1)                                           & \text{(def. $\ftF$)} \\ 
\end{array} 
\]
Finally, $\ftF$ being extensive follows directly from the definition of $\ftF$ and the properties
of $\lub$. \qed

\subsection{Semantics for STE}

Before giving our semantics for STE we first introduce the concept of 
\emph{trajectories}:

\head{Trajectories}
A trajectory is defined as a sequence in which no more information can be
derived by forwards propagation. That is, a sequence $\tau$ is a trajectory   of a
closure function when it is a fixed-point of the closure function for sequences.
%
So, a sequence $\tau$ is a
\emph{trajectory} of $F$ iff
$ 
\tau = \ftF(\tau).
$
%
%

\head{STE-assertions}  have the form $A \Longrightarrow C$.
Here $A$ and $C$ are formulas in  \emph{Trajectory
Evaluation Logic} (TEL). The only variables in the logic are time-independent Boolean
variables taken from the set $V$ of \emph{symbolic constants}.
The language is given by
the following grammar:
\[
f \; ::= \; n \is 0 \; \; | \; \; n   \is 1 \; \; | \; \; f_1 \en f_2 \; \; | \; \; P \rightarrow f \; \; | \; \; \N f \\
\]
\noindent where $n$ is a circuit node and $P$ is a Boolean propositional formula over the set of
symbolic constants $W$.  
%
The operator $\is$ is used to make a statement about the Boolean value of a particular
node in the circuit, $\mathbf{and}$ is conjunction, $\rightarrow$ is used to
make conditional statements, and $\N$ is the next time operator. Note that symbolic
constants only occur in the Boolean propositional expressions on the left-hand side
of an implication.
The notation $n \is P$, where $P$ is a Boolean symbolic expression over the set of
symbolic constants $V$, is used to abbreviate the formula:
$
(\neg P \rightarrow n \is 0) \en (P \rightarrow n \is 1).
$

The \emph{depth} of a TEL-formula $f$ is the maximal degree of nestings of $\N$ in $f$.
The depth of an STE-assertion $A \Longrightarrow C$ is the maximum of the depth of $A$ and the depth of $C$.

The meaning of a TEL formula is defined by a satisfaction relation that  relates
valuations of the symbolic constants and sequences to TEL formulas.  Here, the following
notation is used: The time shifting operator $\sigma^1$ is defined by $\sigma^1(t)(n) =
\sigma(t+1)(n).$ Standard propositional satisfiability is denoted by $\modPropTwo$.
Satisfaction of a trajectory evaluation logic formula $f$ of depth $d$, by a sequence
$\sigma$ of at least depth $d$, and a valuation $\phi : W \rightarrow \{0,1\}$ (written $\phi,\sigma \models f$) is
defined by
 \[
\hspace{-2.2cm}
\begin{array}{lcl}
\phi,\sigma \models  n \is b     & \;\; \equiv \;\; & \sigma(0)(n) = b \: \: \:, \: \: b \in \{0,1\} \\ 
\phi,\sigma \models  f_1 \en f_2 & \equiv &
\phi,\sigma \models  f_1 \;\; \mathrm{and} \;\;  \phi,\sigma \models  f_2 \: \\
\phi,\sigma \models  P \rightarrow f            & \equiv &
\phi \modPropTwo P \;\; \mathrm{implies} \;\; \phi,\sigma \models f \\
\phi,\sigma \models  \N f                       & \equiv & \phi,\sigma^1 \models f\\
\end{array}
\]


\head{Semantics for STE} 
We introduce three semantics for STE. They differ in the way that is dealt
with the over-constrained value $\T$.    
There are several ways of dealing with this 
value 
in a semantics
for STE. 

First of all, we can treat $\T$ as a global contradiction. That is, a sequence that gives value $\T$
to any node, satisfies any antecedent and consequent. So, in order to check whether an
STE-assertion holds we need only consider three-valued sequences. 
\begin{defi}
A circuit  with closure function $F$
satisfies a trajectory assertion $A \Longrightarrow C$ of depth $d$, 
written $F \models A \Longrightarrow C$,  iff for every valuation $\phi : W \rightarrow \{0,1\}$ of the
symbolic constants, and for
every three-valued trajectory $\tau$ of $F$ of depth $d$, it holds that:
\[
\phi,\tau \models A  \;\; \Rightarrow \;\; \phi,\tau \models C.
\]
\end{defi}
Secondly, we can treat $\T$ as a local contradiction. For example, the requirement that 
a node should have value $1$ is fulfilled if the node has value $\T$. But other, unrelated
requirements are unaffected. We introduce 
the \emph{simple semantics for STE} using this approach.
\begin{defi}
A circuit  with closure function $F$ 
\emph{simply satisfies} a trajectory assertion $A \Longrightarrow C$ of depth $d$,
 written $F \modS A \Longrightarrow C$,    iff for every valuation $\phi : W \rightarrow \{0,1\}$ of the
symbolic constants, and for
every trajectory $\tau$ of $F$ of depth $d$, it holds that:
\[
\phi,\tau \models A  \;\; \Rightarrow \;\; \phi,\tau \models C.
\]
\end{defi}
The simple semantics turns out to be useful when we compare the proving power
of STE and GSTE in a precise way in Sect.\ \ref{sec:comparing}. In the simple
semantics, it is for example meaningful to talk about what happens in a
sequence {\em before} certain nodes get a value $\T$ forced by an antecedent.

Finally, we can treat $\T$ as an error. That is, if a node is required to have
value $\T$ by the antecedent of an STE-assertion, the STE-assertion is not true. This is the default approach
taken in Intel's in-house verification toolkit Forte \cite{forte}: it raises an
\emph{antecedent failure} if a node is required to have value $\T$ by the antecedent.
We call this semantics, the \emph{cautious semantics for STE}.
\begin{defi}
A circuit with closure function $F$
\emph{cautiously satisfies} a trajectory assertion $A \Longrightarrow C$ of depth $d$, written $F \modC A \Longrightarrow C$, 
if \emph{both} $F \models A \Longrightarrow C$
\emph{and}
for every valuation $\phi$ of the symbolic constants 
there exists a three-valued trajectory $\tau$ of depth $d$ such that
$
\phi,\tau \models A
$.
\end{defi}


\begin{exa}
For an AND-gate with inputs
$\nin_1$ and $\nin_2$, and output $\nout$,
the assertion
\[
(\nout \is 1) \en (\nin_1 \is a) \en (\nin_2 \is b)  
\Lra  
(\nin_1 \is 1) \en (\nin_2 \is 1) 
\]
is true in the normal semantics but not in the cautious semantics. 

For valuations that give at least one of the symbolic
constants $a$ and $b$ the value $0$, there are no three-valued trajectories that meet the
antecedent: there are no three-valued trajectories in which at least one of the inputs of the AND-gate (nodes
$\nin_1$ and $\nin_2$) has value $0$, while the output (node $\nout$) has value $1$. 
Only
for the valuation that gives both the symbolic constants value $1$, there exists a
three-valued trajectory that satisfies the antecedent. As this trajectory satisfies the consequent as well, the
assertion  is true in the normal semantics.   \qed
\end{exa}

\weg{
\begin{exa}
The reader can verify that assertions (1) and (3), given in the introduction in the paper, are true
in this semantics, but assertions (2) and (4) are false. 

For instance, for assertion (1), a case distinction can be made on the value of
the symbolic variable $a$. If $\phi(a)=0$ then in every trajectory $\tau$ that
satisfies $A$, node $\np$ has value $0$, so by forwards propagation also node $\nout$
has value 0, so the trajectory also satisfies $C$. By similar reasoning it can
be derived that for $\phi(a)=1$, every trajectory that satisfies $A$ also satisfies $C$.

For assertion (2), a counter-example is the trajectory that gives both $\nin_1$ and $\nin_2$
value $1$, and all other nodes value $\X$. \qed
\end{exa}

\head{Cautious Semantics} In our proposed semantics, as well as in the Y-semantics,
an assertion can be true even if for a particular valuation
$\phi$ of the symbolic constants there are no three-valued trajectories of the circuit
that satisfy the antecedent. This is  illustrated in the example below.
\begin{exa}
For an AND-gate with inputs
$\nin_1$ and $\nin_2$, and output $\nout$,
the assertion
\[
(\nout \is 1) \en (\nin_1 \is a) \en (\nin_2 \is b)  
\Lra  
(\nin_1 \is 1) \en (\nin_2 \is 1) 
\]
is true in the closure semantics. For valuations that give at least one of the symbolic
constants $a$ and $b$ the value $0$, there are no three-valued trajectories that meet the
antecedent: there are no three-valued trajectories in which at least one of the inputs of the AND-gate (nodes
$\nin_1$ and $\nin_2$) has value $0$, while the output (node $\nout$) has value $1$. 
Only
for the valuation that gives both the symbolic constants value $1$, there exists a
three-valued trajectory that satisfies the antecedent. As this trajectory satisfies the consequent as well, the
assertion  is true in the closure semantics.   \qed
\end{exa}

%
%
%

%


The default approach taken in Intel's in-house verification toolkit Forte is to demand that for each valuation of the symbolic
constants there exists at least one three-valued trajectory that satisfies the antecedent. 
In order to warn the user of a possible mistake, Forte reports an `antecedent-failure' when
the top-value $\T$ is required to satisfy the antecedent.
We have formalised this approach in the
\emph{cautious closure semantics} of STE. A circuit with closure function $F$ 
{\em cautiously satisfies} a
trajectory assertion $A \Longrightarrow C$ iff \emph{both} $F$ satisfies $A \Longrightarrow C$
\emph{and}
for every valuation $\phi$ of the symbolic constants 
there exists a three-valued trajectory $\tau$ such that
$
\phi,\tau \models A
$.
}

%% file: gste-semantics.tex
\section{A Faithful Semantics for GSTE}

In this section, we present an alternative semantics for GSTE.
As stated in the introduction, there are two reasons for doing
so. 
First of all, the existing semantics for GSTE~\cite{gsteNotPublished} are not faithful to
the proving power of GSTE algorithms. Therefore, they cannot
be used to understand or predict whether certain properties
can be proven by GSTE model checkers.
Secondly, 
a faithful semantics for GSTE can be used 
as basis for research on new GSTE model checking algorithms
and other GSTE tools.

The semantics presented in this section is built up in the same way as the
semantics for STE in the previous section. First, we introduce
the concept of \emph{sequence graphs}. Like sequences in 
STE, sequence graphs represent circuit behaviour over time.

Then, we define a \emph{closure function for sequence graphs}.
Comparable to the closure function for sequences in STE, the
closure function for sequence graphs, given a sequence graph,
calculates all information
that can be derived by forwards propagation of information.

After that, we introduce the concept of \emph{trajectory graphs}.
A trajectory graph is a sequence graph in which no more
information can be derived by forwards propagation of 
information. Thus, a sequence graph is a trajectory graph
precisely when it is a fixpoint of the closure function 
for sequence graphs.

Then, we formally define the concept of \emph{assertion
graphs}. Examples of assertion graphs are given in the introduction of this 
paper. An assertion graph describes a property of the behaviour
of the circuit, possibly ranging over unbounded time.

Finally, by combining all these concepts, we introduce a faithful semantics for GSTE.

\subsection{Sequence Graphs}

\begin{figure}[t]
\[
\epsfig{file=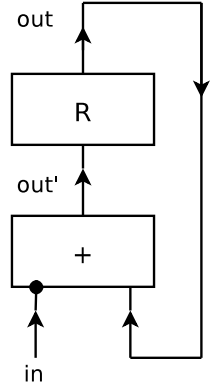,width = 3cm}
\]
\caption{A simple circuit}
\label{fig:circ1}
\end{figure}

We introduce the concept of \emph{sequence graphs} to represent circuit 
behaviours over time. They are comparable to the concept of sequences in STE.
Sequence graphs, however, are more expressive: each sequence graph represents a 
(possibly infinite) number of sequences.
\begin{exa}
Consider the circuit given in Figure \ref{fig:circ1}.
The following picture represents a sequence graph of the circuit. 
\begin{equation}
\xymatrix{ 
{\init} \ar@/^/[r]^{11\X} \ar@/_/[r]_{0\X\X}   & v \ar[r]^{\X\X\X}  & w \ar@(ur,dr)^{000}
}
\label{sg1}
\end{equation}
The sequence graph has vertices $\init,v$ and $w$, two edges from $\init$ to $v$, an edge
from $v$ to $w$, and an edge  from $w$ to itself. In the picture, states are represented
by vectors of truth-values, in the order $\nin,\noutp$,$\nout$.  For instance, in the state
represented by $11\X$, node $\nin$ has value $1$, node $\noutp$ has value $1$, and node
$\nout$ has value $\X$.

Each  \emph{path} in the graph starting in initial vertex $\init$, represents a possible
behaviour of the circuit over time. For instance, consider the path starting in $\init$, going through
the top edge between $\init$ and $v$, and then cycles twice through the looping
edge at vertex $w$. This path represents the sequence 
$$
[11\X,\X\X\X,000,000]\eqno{\qEd}
$$
\end{exa}
The reader should note the difference between \emph{sequence graphs} and
\emph{assertion graphs} (see page \pageref{ag_mc1} for an example of an
assertion graph). Sequence graphs represent \emph{circuit behaviour}
(corresponding to {\em sequences} in STE), whereas assertion graphs describe
\emph{desired properties} of circuit behaviour (corresponding to {\em
assertions} in STE).

\begin{defi}
A \emph{sequence graph} is a triple $(V,E,\Sigma)$, where:
\begin{enumerate}[$\bullet$]
\item{$V$ is a finite set of vertices containing the \emph{initial vertex} $\init$.}
\item{$E$ is a finite set of directed edges between vertices. Each edge $e$ has
a \emph{start vertex} $\vstart(e)$ and an \emph{end vertex} $\vend(e)$. 
Multiple edges between two vertices are allowed.
}
\item{$\Sigma : E \rightarrow \CircuitState$ is a function from edges to circuit states.}  

\end{enumerate}
We say that sequence graphs $(V_1,E_1,\Sigma_1)$ and $(V_2,E_2,\Sigma_2)$ 
are of the \emph{same shape} iff $V_1=V_2$ and $E_1=E_2$. The set of all sequence graphs is denoted $\SeqGraph$.
\end{defi}
Usually, a sequence graph is identified by the function $\Sigma$ only.

The order $\leq$ and the operators $\lub$ and $\glb$ on the domain $\{0,1,\X,\T\}$ are extended
in a point-wise fashion to pairs of sequence graphs of the same shape.
That is, the order $\leq$ on sequence graphs is defined by $\Sigma_1 \leq \Sigma_2$ iff for all
edges $e$ and nodes $n$, $\Sigma_1(e)(n) \leq \Sigma_2(e)(n)$. 
Furthermore, $(\Sigma_1 \lub \Sigma_2)(e)(n)=(\sigma_1(e)(n) \lub \sigma_2(e)(n))$
and $(\Sigma_1 \glb \Sigma_2)(n)=(\Sigma_1(e)(n) \glb \Sigma_2(e)(n))$.

An edge is \emph{initial} if it starts in the initial vertex $\init$.
We define the set of incoming edges of an edge $e$, written $\ine(e)$ by:
\[
\ine(e) = \{ \: e' \in E \: | \: \vstart(e)=\vend(e') \}  
\]

A \emph{path of depth $d$} is a list of edges $\rho = (e_0,e_1,\ldots,e_d)$ such that
for each $i$, $\vstart(e_{i+1})=\vend(e_{i})$. 
An \emph{initial path} is a path whose first edge is initial.

A finite initial path $\rho$ of depth $d$ in a sequence graph $\Sigma$ represents the sequence $\seqp(\Sigma,\rho)$
of depth $d$ defined by
\[
\seqp(\Sigma,\rho)(t) = \Sigma(\rho(t)). 
\]
A sequence graph $\Sigma$ represents the set of sequences $\seqp(\Sigma)$ defined by
\[
\seqp(\Sigma) 
= \{\seqp(\Sigma,\rho) \: | \: \rho \text{ is a finite initial path in } \Sigma \} 
\]
We will only consider sequence graphs in which each edge and each vertex is reachable from the
initial vertex $\init$. That is, we require that for each edge there exists an initial path containing the edge,
and for each vertex there exists an initial path containing the vertex.
The reason is that states at unreachable edges cannot appear in the sequences
represented by the sequence graph.
\begin{exa}
The sequence graph (\ref{sg1}) represents the following infinite set of sequences:

\[
\begin{array}{l}
[11\X] \\{}
[11\X,\X\X\X] \\{}
[11\X,\X\X\X,000] \\{}
[11\X,\X\X\X,000,000] \\{}
\ldots \\{}
[0\X\X] \\{}
[0\X\X,\X\X\X] \\{}
[0\X\X,\X\X\X,000] \\{}
[0\X\X,\X\X\X,000,000] \\{}
\ldots \\{}
\end{array}
\]
%
%
The sequence graph
\[
\xymatrix{ 
{\init}  \ar[r]^{111} & v \ar@(ul,ur)^{\X\X\X} \ar[r]^{000} &  w    
}
\]
represents the following infinite set of sequences:
\[
\begin{array}{l}
[111] \\{}
[111,000] \\{}
[111,\X\X\X] \\{}
[111,\X\X\X,000] \\{}
[111,\X\X\X,\X\X\X] \\{}
[111,\X\X\X,\X\X\X,000] \\{}
\ldots
\end{array}
\]\vskip-16 pt
\qed
\end{exa}

\subsection{Trajectory Graphs and Closure Functions}

We introduce the concept of \emph{trajectory graphs} to represent
sequence graphs in which no more information can be derived by
forwards propagation of information.  Trajectory graphs are comparable
to the concept of trajectories in STE.

In order to define trajectory graphs, we define a \emph{closure function
for sequence graphs}, written $\Fsg : \SeqGraph \rightarrow \SeqGraph$. 
The idea is that such a closure function, given a sequence graph, derives
all information that can be derived by forwards propagation. Then, a trajectory
graph is defined as a fixpoint of this closure function.
 
Before doing so, let us first get some more intuition on desired
properties for a closure function for sequence graphs. Recall that a
sequence graph represents a (possibly infinite) collection of
sequences. Each initial path $\rho$ in the graph represents a sequence
$\seqp(\Sigma,\rho)$ as defined before.

Furthermore, recall, from the introduction, when a circuit satisfies a
GSTE assertion graph, the circuit should also satisfy all
STE-assertion corresponding to finite initial paths in the assertion
graphs.

Therefore, given a sequence graph and an initial path $\rho$ in the
graph, we expect that the closure function on sequence graphs $\Fsg$
for the edges in $\rho$ derives at most the information as the closure
function for sequences $\Fseq$ does for the sequence
$\seqp(\Sigma,\rho)$. The reason for requiring this is that if the
closure function on sequence graphs were to derive more information
for a particular sequence in the sequence graph than the closure
function on sequences, then we could construct a GSTE assertion graph
that is satisfied by the circuit, but that contains a path
corresponding to an untrue STE-assertion.

So, we require the following property:
\begin{pty} 
A closure function $\Fsg$ for sequence graphs
\emph{derives no more information} than a closure function on sequences $\Fseq$, if 
for all sequence graphs $\Sigma$ and initial paths $\rho$,
\[
\seqp(\Fsg(\Sigma),\rho) \leq \Fseq (\seqp(\Sigma,\rho))
\]
\label{prop:Fsg}
\end{pty}
\vskip-12 pt

The closure function for sequence graphs is allowed to
derive \emph{less} information for a particular path than the closure
function for sequences does. The reason is that an edge might be
reached via different initial paths. If, for these paths, the
closure function for sequences derives conflicting values
for a circuit node at that edge, the above property forces
the circuit node to take on value $\X$. We elaborate on this
in Example \ref{ex:nonlin2} on page \pageref{ex:nonlin2}.

Defining a closure function for sequence graphs is a greater challenge
than defining a closure function for sequences. There are two reasons
for this.

First of all, in STE, for a state at time $t+1$, there is precisely
one ``previous" state, namely the state at time $t$. So, it is clear
how the information from previous points in time should be
propagated. In GSTE, however, a state at an edge can have multiple
predecessors. So, we have to decide how to combine information from
incoming edges.

Secondly, in STE-sequences, the state at each time-point depends only on the
previous states, and, thus, never on itself. In GSTE sequence graphs, however,
cycles may be present, therefore  a state may, via a cycle, depend on itself. 

In the following, we gradually construct a closure function for sequence 
graphs by considering closure functions for increasingly more complex
sets of sequence graphs. First, we only consider \emph{linear} sequence graphs,
then we look at \emph{acyclic} sequence graphs, and finally we consider
general sequence graphs.

\subsubsection*{Linear sequence graphs}
Let us take one step at the time. So, first, assume we have a sequence graph
where each vertex has at most one successor and at most one predecessor, and
no cycles are present. Such a sequence graph has the following form:
\[
\xymatrix{ 
{\init}  \ar[r] & v_1 \ar[r] & v_2  \ar[r] & {\ldots} \ar[r] & v_d   
}
\]
For edges that have exactly one incoming edge, we define the function $\pre$
$\{\pre(e)\} = \ine(e)$. (Recall that $\ine(e)$ is the set of all incoming
edges of $e$.)

For the above sequence graph, a closure function $\FsgLine$ can be defined in
the same way as in STE. For the initial edge, no information is propagated
from a previous time-point, so only closure of the initial state is needed. 
For each other time-point, information from the previous state should be propagated.

This yields the closure function $\FsgLine$:
\[
\FsgLine(\Sigma)(e) = 
\twocases
{F(\Sigma(e))}{\text{$e$ is initial}}
{F(\Sigma(e) \tlub \nextf( \: \FsgLine(\Sigma)(\pre(e)) \:)}{\text{otherwise}}
\]
The function $\FsgLine$ is well defined as the graphs we consider here are
acyclic and edges have at most one predecessor. Note the similarity
with the closure function for sequences $\Fseq$ on page \pageref{def:Fseq}.

Note that, just like in STE, the initial values of registers are ignored.

$\FsgLine$ calculates precisely the same information as $\Fseq$.
That is, for each initial path $\rho$ in a sequence graph $\Sigma$ of the above
form,
\[
\seqp(\FsgLine(\Sigma),\rho) = \Fseq (\seqp(\Sigma,\rho)).
\]

\subsubsection*{Acyclic sequence graphs}

Now, let us consider a more general situation:
an acyclic graph.

\begin{exa}
Consider the following sequence graph: 
\[
\xymatrix{ 
{\init} \ar@/^/[r]^{11\X} \ar@/_/[r]_{000}   & v \ar[r]^{\X\X\X}   &  w
}
\]
The edge starting at $v$ has two incoming edges. 
For this edge, the state at the
previous time-point can be any of the states at the predecessor edges,
that is $11\X$ and $000$.
The first state gives node $\noutp$ value $1$, so if this state
had been the only predecessor state, we would have concluded
that node $\nout$ should have value $1$ at the edge starting at $v$. 
However, the state at the second incoming edge gives node $\noutp$
value $0$, so according to this state, node $\nout$ should have value $1$ at the edge starting at $v$. 
Therefore, as the two incoming edges do not agree on the
value of node $\nout'$,
nothing can be derived about the value of node 
$\nout$ at the edge starting in $v$.
So, no more information can be derived from this sequence graph. \qed
\label{ex:nonlin}
\end{exa}
So, only if all the states of the incoming edges agree on a Boolean value of an input to a
register,  should this value be propagated to the output of the register. Thus,
to combine the values on the inputs of the register, the \emph{greatest lower bound}
should be used.
This yields the closure function $\FsgNoCycle$:
\[
\FsgNoCycle(\Sigma)(e) = 
\twocases
{F(\Sigma(e))}{\text{$e$ is initial}}
{F(\Sigma(e) \tlub \mathop{\glb}\limits_{i \in \ine{(e)}} \nextf( \: \FsgNoCycle(\Sigma)(i)) \:)}{\text{otherwise}}
\]
\begin{exa}
Applied to the sequence graph in Example \ref{ex:nonlin}, the closure function yields the same sequence graph.
%
In the graph, the top edge between $\init$ and
$v$ gives value $1$ to $out'$, the bottom edge gives value $0$ to
this node, so $0 \tglb 1 = \X$ is propagated for the value of node
$\nout$ for the edge starting in $v$.

In this example, the closure function for sequence graphs derives
less information than the closure function for sequences for the paths
in the assertions. The reason is that the greatest lower-bound operator
is used to combine conflicting information from incoming edges.

Applied to
\[
\xymatrix{ 
{\init} \ar@/^/[r]^{1\X\X} \ar@/_/[r]_{1\X\X}   & v \ar[r]^{\X\X\X}  & w
}
\]
the closure function yields
\[
\xymatrix{ 
{\init} \ar@/^/[r]^{11\X} \ar@/_/[r]_{11\X}   & v \ar[r]^{\X11}  & w
}
\]
As both incoming edges give value $1$ to node $\noutp$ this value is propagated
to node $\nout$. \qed
\label{ex:nonlin2}
\end{exa}

\subsubsection*{General sequence graphs}

Now that we have dealt with sequence graphs where edges can have multiple
predecessors, it is time to tackle the next challenge: cycles.
The following example illustrates that when cycles are present,
the equations for $\FsgNoCycle$ no longer define a function, but,
instead, may have more than one solution.

\begin{exa}
Consider the sequence graph:
\begin{equation}
\xymatrix{ 
{\init}  \ar[r]^{1\X\X} & v \ar@(ur,dr)^{\X\X\X}    
}
\label{sgCycle}
\end{equation}
%
Here, the result for the initial edge still can be calculated (yielding $11\X$),
but the result for the self-loop at vertex $v$ is problematic. 
The equations state:
\[
\FsgNoCycle(\Sigma)((v,v)) = F ( \X\X\X \tlub ( \nextf(11\X) \tglb
\nextf(\FsgNoCycle(\Sigma)((v,v))) ))  
\]
This can be simplified to:
\[
\FsgNoCycle(\Sigma)((v,v)) = F (  \X11 \tglb \nextf(\FsgNoCycle(\Sigma)((v,v))) ),  
\]
further simplified to:
\[
\FsgNoCycle(\Sigma)((v,v)) = F (  \X11 \tglb \X\X(\FsgNoCycle(\Sigma)((v,v))(\noutp)) ),  
\]
and finally simplified to:
\[
\FsgNoCycle(\Sigma)((v,v)) = F (  \X\X(1 \tglb \FsgNoCycle(\Sigma)((v,v))(\noutp)) ).  
\]
This equation can be rewritten to:
\[
\FsgNoCycle(\Sigma)((v,v)) = (\lambda s . F (\X\X(1 \tglb s((v,v))(\noutp)) ) \: \: \FsgNoCycle(\Sigma)((v,v))  
\]
This equation has as solutions precisely the fixpoints of
\[
(\lambda s . F (\X\X(1 \tglb s(\noutp)) )
\]
The two fixed-points are $\X\X\X$ or $\X11$.

The first fixed-point yields the following sequence graph:
\begin{equation}
\xymatrix{ 
{\init}  \ar[r]^{11\X} & v \ar@(ur,dr)^{\X\X\X}    
}
\label{sgCycleFix1}
\end{equation}
This contradicts our intuition: if at the first point in time 
node $\nin$ has value $1$, then we expect that, from the next time-point on,
node $\nout$ and $\noutp$ have value $1$ as well. So, only
the second fixed-point gives the expected sequence graph:
\begin{equation}
\xymatrix{ 
{\init}  \ar[r]^{11\X} & v \ar@(ur,dr)^{\X11}    
}
\label{sgCycleFix2}
\end{equation}
\qed
\end{exa}\nobreak
So, in general, when cycles are introduced, the equations for
$\FsgNoCycle$ no longer define a function: the equations may have more
than one solution. Let us study this set of solutions more closely.

To do so, we define, for a given sequence graph $\Sigma$, the function
$\FsgTwo : \SeqGraph \rightarrow \SeqGraph$ by:
\[
\FsgTwo(\Delta)(e) = 
\twocases
{F(\Sigma(e))}{\text{$e$ is initial}}
{F(\; \Sigma(e) \tlub \mathop{\glb}\limits_{i \in \ine(e)} \nextf(\Delta(i)) \;)}{\text{otherwise}} 
\]

\noindent Using this, the equations for $\FsgNoCycle$ can be rewritten to:
\[
\FsgNoCycle (\Sigma) =  \FsgTwo(\FsgNoCycle(\Sigma))
\]
The solutions of this equation are the set of fixpoints of $\FsgTwo$.
For example, for $\Sigma$ equal to sequence graph (\ref{sgCycle})
the fixpoints are sequence graphs (\ref{sgCycleFix1}) and (\ref{sgCycleFix2}).

The following lemma states that each fixpoint satisfies
Property \ref{prop:Fsg}.

\begin{lem}
For each $\Delta$ that is a fixpoint of $\FsgTwo$ and
for each initial path $\rho$ in the sequence graph $\Sigma$, it holds that:
\[
\seqp(\Delta,\rho) \leq \Fseq (\seqp(\Sigma,\rho))
\]
\end{lem}

\proof
The proof is by induction on the position in the sequence. 
The base-case $(t=0)$ follows directly from the definitions of $\seqp$ and $\Fseq$.
The induction hypothesis is:
\[
\seqp(\Delta,\rho)(t) \leq (\Fseq (\seqp(\Sigma,\rho)))(t)
\]
If $\rho(t+1)$ is not initial, then
\[
\seqp(\Delta,\rho)(t+1) = F(\; \Sigma(\rho(t+1)) \tlub \mathop{\glb}\limits_{i \in \ine(\rho(t+1))} \nextf(\Delta(i)) \;)
\]
Now:
\[
\begin{array}{cll}
      & \mathop{\glb}\limits_{i \in \ine(\rho(t+1))} \nextf(\Delta(i)) & \\ 
 \leq & \nextf(\Delta(\rho(t)))                                        & (\rho(t) \in \ine(\rho(t+1))\\
 =    & \nextf(\seqp(\Delta,\rho)(t)))                                 & (\text{Definition } \seqp) \\
 \leq & \nextf(\Fseq (\seqp(\Sigma,\rho))(t))                          & (\text{(Induction hypothesis and monotonicity $\nextf$)} \\
\end{array}
\]
Thus, 
\[
\seqp(\Delta,\rho)(t+1) \leq F(\; \seqp(\Sigma,\rho)(t+1) \tlub
\nextf(\Fseq(\seqp(\Sigma,\rho))(t)) \;) 
\]
So, by the definition of $\Fseq$,
\[
\seqp(\Delta,\rho)(t+1) \leq \Fseq (\seqp(\Sigma,\rho))(t+1)
\]
The case for $\rho(t+1)$ is initial is similar but easier. \qed
%
%
%
%
%
The question now is: which fixpoint should the closure function
for sequence graphs choose? As each fixpoint satisfies Property \ref{prop:Fsg},
it is sound to choose any of them. 
The following property states there exists a unique
greatest fixpoint.

\begin{prop}
For each $\Sigma$, the function $\FsgTwo$
has a unique greatest fixpoint. 
\end{prop}

\proof
It is easy to see that $\FsgTwo$ is monotonic.
The collection of sequence graphs with the same
vertices and edges as $\Sigma$ and giving values to the
same circuit nodes as $\Sigma$ is finite and forms, together
with the order $\leq$ on sequence graphs,
a complete lattice. So,
by 
Tarski's fixpoint theorem~\cite{tarski}, $\FsgTwo$
has a greatest fixpoint. \qed
As each fixpoint satisfies Property \ref{prop:Fsg}, we can safely choose the greatest
one, giving the most information.
Thus, we define the  closure function for sequence graphs as follows: 

\begin{defi} Given a closure function $F$, the \emph{closure function for sequence
graphs}, written $\Fsg : \SeqGraph \rightarrow \SeqGraph$ is defined by:
\[
\Fsg(\Sigma) = \gfp \Delta . \FsgTwo(\Delta)
\]
\end{defi}
%
%

\begin{prop}
Given a closure function $F$, $\Fsg$ is a closure function as well.
\label{prop:FsgCF}
\end{prop}

\proof
Suppose $F$ is a closure function, we have to prove that $\Fsg$ is monotonic, extensive and idempotent.
$\Fsg$ being extensive follows directly from the definition of $\Fsg$.

We now prove that $\Fsg$ is monotonic.
Suppose $\Sigma_1 \leq \Sigma_2$, $\Delta_1 = \Fsg(\Sigma_1)$ and $\Delta_2 =
\Fsg(\Sigma_2)$, then
\[
\Delta_1 = \Fsg_{\Sigma_1}(\Delta_1) \leq \Fsg_{\Sigma_2}(\Delta_1) 
\]
Tarski's fixpoint theorem~\cite{tarski} states that 
\[
\gfp \Delta . \Fsg_{\Sigma_2}(\Delta) = \lub \{\Delta \: | \: \Delta \leq \Fsg_{\Sigma_2}(\Delta)   \}
\]
Thus $\Delta_1 \leq \gfp \Delta . \Fsg_{\Sigma_2}(\Delta) = \Delta_2$.

Finally, we prove that $\Fsg$ is idempotent.
Suppose $\Fsg(\Sigma)=\Delta$ and $\Fsg(\Delta)=\Delta'$.
We need to prove that $\Delta = \Delta'$. 
By monotonicity of $\Fsg$ follows $\Delta \leq \Delta'$.
We prove that $\Delta' \leq \Delta$ by proving that
$\Delta'$ is a fixpoint of $\FsgTwo$ (then, because $\Delta$ is 
the greatest fix-point of $\FsgTwo$, it follows that $\Delta'\leq\Delta$). The case for
when $e$ is initial is trivial. Suppose $e$ is not initial.
\[
\begin{array}{cll}
      & \FsgTwo(\Delta')(e) & \\ 
 =    & F(\Sigma(e) \tlub \mathop{\glb}\limits_{i \in \ine(e)} \nextf(\Delta'(i))) & (\text{Definition } \FsgTwo )\\
 =    & F(\Sigma(e) \tlub \mathop{\glb}\limits_{i \in \ine(e)} \nextf(\Delta(i)) \tlub \mathop{\glb}\limits_{i \in \ine(e)}
              \nextf(\Delta' (i)) ) & ( \text{Prop } \lub, \text{ } \Delta \leq \Delta' ) \\
 =    & F(F(\Sigma(e) \tlub \mathop{\glb}\limits_{i \in \ine(e)} \nextf(\Delta(i))) \: \: \tlub \mathop{\glb}\limits_{i \in \ine(e)}
              \nextf(\Delta' (i)) ) & ( F \text{ is closure function} ) \\
 =    & F( \Delta(e) \tlub \mathop{\glb}\limits_{i \in \ine(e)}
              \nextf(\Delta' (i)) ) & ( \Delta \text{ is fixpoint of } \FsgTwo ) \\
 =    & \Delta'(e)                 & ( \Delta' \text{ is fixpoint of } \Fsg_{\Delta} )
\end{array}
\]
\qed
%
%

\subsubsection{Trajectory Graphs}
We define a \emph{trajectory graph} of $F$ as a sequence graph that is a fixpoint
of $\Fsg$. 

\begin{defi}
A sequence graph $\Sigma$ is a \emph{trajectory graph} of a closure function $F$, if
\[
\Fsg(\Sigma) = \Sigma
\]
\end{defi}

%% file: gste-semantics2.tex
\subsection{Assertion Graphs}

In GSTE, circuit properties are given by \emph{assertion graphs}.
An example of an assertion graph is:
\begin{equation}
\xymatrix{ 
{\init}  \ar[r]^{\nin \is 1 / \True} & v \ar@(ur,dr)^{\True/\nout \is 1}    
}
\label{ag1}
\end{equation}
 
\noindent
In the assertion graph, each edge is labelled with a pair $A/C$, here
$A$ is called the antecedent and $C$ is called the consequent. Just like in 
STE, the antecedent represents assumptions made, and the consequent 
represents requirements.

Both $A$ and $C$ are, like in STE, formulas in trajectory evaluation logic ($\TEL$). However, as
each edge represents the state of a single time-point, no occurrences of the next-time 
operator $\N$ are allowed. We call the subset of $\TEL$ in which no next-time operators occur $\GTEL$.

The assertion graph above states that if at some time point, node $\nin$
has value 1, then at each later time-point node $\nout$ has
value 1 as well. 
\begin{defi}
An assertion graph is a four-tuple $G=(V,E,\ant,\cons)$.
Here, $V$ is a set of \emph{vertices} containing a vertex $\init$ which is called the \emph{initial vertex},
$E$ is a set of \emph{edges} between the vertices. 
Finally, $\ant,\cons : E \rightarrow \GTEL$ are functions from edges to 
formulas in $\GTEL$.
\end{defi}
Recall that path is called initial iff it starts in the initial
vertex $\init$.
A finite initial path $\rho$ of depth $d$ in an assertion graph $G$ 
represents an STE assertion  $\ass(G,\rho)$ defined by
\[
\ass(G,\rho) = 
(\mathop{\en}\limits_{0 \leq i \leq d} \N^i \ant(\rho(t))) 
\Rightarrow 
(\mathop{\en}\limits_{0 \leq i \leq d} \N^i \cons(\rho(t))) 
\]
An assertion graph represents a (possibly infinite) collection
of STE-assertions: for each finite initial path $\rho$ in the graph,
an STE-assertion $\ass(G,\rho)$. The set of STE-assertions
in assertions graph $G$, written $\ass(G)$, is defined by:
\[
\ass(G) = \{\ass(G,\rho) \: | \: \rho \text{ is a finite initial path in } G\}
\]
\begin{exa} Assertion graph (\ref{ag1}) above
represents the following infinite set of STE-assertions:
\[
\begin{array}{lcl}
\nin \is 1 & \Longrightarrow & \N(\nout \is 1)  \\
\nin \is 1 & \Longrightarrow &  \N(\nout \is 1) \en \N\N(\nout \is 1)  \\
\nin \is 1 & \Longrightarrow & \N(\nout \is 1) \en \N\N(\nout \is 1) \en \N\N\N(\nout \is 1)  \\
\multicolumn{3}{c}{\ldots} \\
\end{array}
\]\vskip-16 pt
\qed
\end{exa}

The idea is that when a circuit satisfies a GSTE assertion 
graph, the circuit graph also satisfies all STE assertions
in the assertion graph. The converse, however, does not hold,
as we will see in the next section.
\vfill\eject

\subsection{Satisfiability}

Satisfaction of a $\GTEL$-formula $f$, by a circuit state
$s : \CircuitState$ and a valuation $\phi : W \rightarrow \{0,1\}$ of
the symbolic constants
(written $\phi,s \models f$) is
defined by
\[
\begin{array}{lcl}
\phi,s \models  n \is b     & \;\; \equiv \;\; & \sigma(0)(n) = b \: \: \:, \: \: b \in \{0,1\} \\ 
\phi,s \models  f_1 \en f_2 & \equiv &
\phi,s \models f_1 \;\; \mathrm{and} \;\;  \phi,s \models  f_2 \: \\
\phi,s \models  P \rightarrow f            & \equiv &
\phi \modPropTwo P \;\; \mathrm{implies} \;\; \phi,s \models f \\
\end{array}
\]
\begin{exa}
If $s(\nin)=1$ and $s(\nout)=0$, and $\phi(a)=1$ and $\phi(b)=0$, then 
$$
\phi,s \models (\nin \is a) \en (\nout \is b) \en (\nin \is \neg(a \wedge b))
\eqno{\qEd}
$$
\end{exa}
We say that a sequence graph $(V,E,\Sigma)$ satisfies a function $f : E \rightarrow \GTEL$,  
$f \in \{\ant,\cons\}$ and 
a valuation $\phi : W \rightarrow \{0,1\}$ of
the symbolic constants,
written $\phi,\Sigma \models f$, if for all edges $e$:
\[
\phi,\Sigma(e) \models f(e)
\]
Note that the definition of satisfaction above requires that the shape of the sequence
graph be identical to the shape of the assertion graph from which the antecedent or consequent
is taken.
\begin{exa}
If $G=(V,E,\ant,\cons)$ is assertion graph (\ref{ag1}), $\Sigma_1$ is 
sequence graph $(\ref{sgCycle})$, and
$\Sigma_2$ is sequence graph $(\ref{sgCycleFix2})$,
then for any $\phi$: 
$\phi,\Sigma_1 \models \ant$,
$\phi,\Sigma_2 \models \ant$,
$\phi,\Sigma_1 \not\models \cons$,
and
$\phi,\Sigma_2 \models \cons$. \qed
\end{exa}
Just like in STE, in GSTE, there are several ways of dealing with 
the over-constrained value $\T$. We can treat $\T$ just as any other
value, leading to the \emph{simple semantics} of GSTE. Or, we 
can treat an over-constrained value as an error, leading to the
\emph{cautious semantics} of GSTE.

In GSTE, however, we cannot treat $\T$ as a contradiction in
the same way as we did in STE. The reason is the following.
Consider a semantics
in which a sequence graph that assigns a $\T$ to a circuit node at an edge
satisfies any antecedent and consequent.
In such a semantics, GSTE assertion graphs 
containing false STE-assertions may still be true. For example, given a GSTE
assertion graph containing a false STE assertion, we can 
simply add a fresh initial edge with an inconsistent antecedent,
making the GSTE assertion true. 

If, instead,  we require a $\T$ value at \emph{each} path in the graph
to deem a sequence graph contradictory, this problem does not occur. However,
as the implementation of such a semantics in a GSTE model checker
seems cumbersome, we will not elaborate on such a semantics
further.

In the definition of \emph{simple satisfaction} for GSTE, 
the value $\T$ is treated just like any other value, and models
a local conflict of demands made by the assertion. 
In this paper, we consider this the `standard' semantics for GSTE. As explained
in Sect.\ \ref{sec:comparing}, this turns out to correspond well with what
most GSTE algorithms do in practice.
%
%
\begin{defi}
We say that a closure function $F$ \emph{simply satisfies} an assertion graph \linebreak $G=(V,E,\ant,\cons)$, 
written $F \modS G$,
if for all assignments of symbolic constants $\phi : W \rightarrow \{0,1\}$, trajectory
graphs $\Sigma$, 
\[
\Sigma \models \ant \Rightarrow \Sigma \models \cons.
\]
\end{defi}

\begin{exa}
If $G=(V,E,\ant,\cons)$ is assertion graph (\ref{ag1}),  and $F$ is the closure function for the circuit in Figure
\ref{fig:circ1}, then $F \modS G$. 

This can be explained as follows. It is easy to see that, for any $\phi$, sequence graph 
$(\ref{sgCycle})$ is the  weakest sequence graph that makes the antecedent of $G$
true.  Let us call this sequence graph $\Sigma$. Trajectory graph  $(\ref{sgCycleFix2})$ is
$\Fsg(\Sigma)$. We claim that $\Fsg(\Sigma)$ is the weakest trajectory graph satisfying
$\ant$.  This can be proven easily. Suppose $T$ is a trajectory graph satisfying $\ant$, then
$\Sigma \leq T$, so by monotonicity of $\Fsg$ and because $T$ is a fix-point of $\Fsg$, 
$\Fsg(\Sigma) \leq \Fsg(T)=T$. Thus, as the weakest trajectory graph satisfying $\ant$ also
satisfies $\cons$, all trajectory graphs that satisfy $\ant$ satisfy $\cons$ as well. 
So, $F \modS G$.

In Section \ref{sec:mc}, we explain that, in the general case, to check whether a circuit 
simply satisfies an assertion graph, we only have to, for each $\phi$, consider the weakest trajectory graph that 
satisfies the antecedent $\ant$. \qed
\end{exa}
In the definition of cautious satisfaction for GSTE, the value $\T$ is treated as an error.
\begin{defi}
We say that a circuit model $F$, \emph{cautiously satisfies} an assertion graph \linebreak $G=(V,E,\ant,\cons)$, 
written $F \modC G$, if $F$ simply satisfies $G$ and 
for all assignments of symbolic constants $\phi : W \rightarrow \{0,1\}$, there
exists a trajectory graph $\Sigma$ of $F$ such that $\Sigma \models \ant$.
\end{defi}
The following example illustrates the difference between the two definitions.
\begin{exa}
The circuit in Figure \ref{fig:circ1} simply satisfies the following assertion graph. It does,
however, not cautiously satisfy it.
$$
\xymatrix{ 
{\init}  \ar[r]^{\nin \is 1 / \True} & v \ar@(ur,dr)^{ \nout \is 0 / \nout \is 1}    
}\eqno{\qEd}
$$
\end{exa}

%% file: comparing.tex
\section{Comparing with STE}
\label{sec:comparing}

In this section we compare STE with GSTE.
The purpose is to make the relationship between STE
and GSTE model checking clear.

The following proposition states that if a closure function satisfies an
assertion graph, it simply satisfies all STE-assertions in the assertion graph
as well.

\begin{proposition} Given an assertion graph $G=(V,E,\ant,\cons)$ for a circuit with closure function $F$:
\[
F \models G \: \Rightarrow \: ( \text{for all assertions } (A \Longrightarrow C) \in \Ass(G) : 
F \modS (A \Longrightarrow C) )
\]
\label{prop:compare}
\end{proposition}

\proof
Suppose $F \models G$, $\rho$ is a finite path of depth $d$ in $G$,
$A \Longrightarrow C = \Ass(G,\rho)$, $\phi$ a valuation of the symbolic constants, and
$\tau$ a trajectory of $F$ of depth $d$  such that $\phi,\tau \models A$. We need to prove that $\phi,\tau \models C$.

Let $\Sigma$ be the sequence graph that has the same shape as assertion graph $G$
and is further defined by:
\[
\Sigma(e) = \mathop{\glb}\limits_{0 \leq t \leq d, \rho(t)=e} \tau(t) 
\] 
Note that $\Sigma(e)(n)=\T$ for edges not in the path $\rho$.
We now prove that $\phi,\Sigma \models \ant$. As $\tau \models A$, and
$A = \mathop{\en}\limits_{0 \leq t \leq d} \N^t \ant(\rho(t)))$,
for each $t$ holds:
\[
\phi,\tau(t) \models \ant(\rho(t))
\]
Thus for all $e \in E$:
\[
\phi, \mathop{\glb}\limits_{t \in \Nat, \rho(t)=e} \tau(t) \models \ant(e) 
\]
Thus, $\phi,\Sigma \models \ant$. As $\Fsg$ is extensive, 
$\phi,\Fsg(\Sigma) \models \ant$ as well. As $\Fsg(\Sigma)$ is a trajectory
graph, and $F \models G$, it holds that $\phi,\Fsg(\Sigma) \models \cons$.
By Property \ref{prop:Fsg}:
\[
\seqp(\Fsg(\Sigma),\rho) \leq \Fseq (\seqp(\Sigma,\rho))
\]
Thus:
\[
\seqp(\Fsg(\Sigma),\rho) \leq \Fseq (\tau) = \tau
\]
Now, as $\Fsg(\Sigma) \models \cons$, for all $e \in E$:
\[
\phi, \Sigma(e) \models \cons(e)
\]
Thus:
\[
\phi, \mathop{\glb}\limits_{0 \leq t \leq d, \rho(t)=e} \tau(t) \models \cons(\rho(t)) 
\]
As $C = \mathop{\en}\limits_{0 \leq i \leq d} \N^i \cons(\rho(t)))$, it follows 
that $\phi,\tau \models C$. \qed

The converse
\[
(\text{for all assertions } (A \Longrightarrow C) \in \Ass(G) : F \modS (A \Rightarrow C)) \: \Rightarrow \: F \models G
\]
however, is not true. The reason is that GSTE combines
conflicting information between incoming edges by using the greatest lower bound
operator. This is illustrated by the following example.

\begin{exa}
Consider the following circuit:
\[
\epsfig{file=circ3.eps,width = 2cm}
\]
The induced closure function of this circuit satisfies the
STE-assertions $\npp \is 1 \Longrightarrow \N (\nout \is 1)$
and  $\npp \is 0 \Longrightarrow \N (\nout \is 1)$. 
Consider the following sequence graph.
In the picture, states are represented
by vectors of truth-values, in the order $\npp,\np,\nout$.
\[
\xymatrix{ 
{\init} \ar@/^/[r]^{1\X\X} \ar@/_/[r]_{0\X\X}   & v \ar[r]^{\X\X\X} & w  
}
\]
The sequence graph is a trajectory graph of the closure function. 
Thus, the closure function does not satisfy the below 
GSTE assertion graph.
$$
\xymatrix{ 
{\init} \ar@/^/[rr]^{\npp \is 1 / \True} \ar@/_/[rr]_{\npp \is 0 / \True}   
&& v \ar[rr]^{\True / \nout \is 1}   
&& w
}\eqno{\qEd}
$$
\end{exa}
The following example shows that if a GSTE assertion graph
is cautiously satisfied (that is, no node has to assume value $\T$
to satisfy the antecedent), there may still be an STE-assertion
represented by the assertion graph that is not cautiously satisfied.
\begin{exa}
Consider a circuit consisting of a single register with 
input $\nregp$ and output $\nreg$, and the following assertion graph.
\[
\xymatrix{ 
{\init} \ar@/^/[rr]^{\nregp \is 1 / \True} \ar@/_/[rr]_{\nregp \is 0 / \True}   
&& v \ar[rr]^{\nreg \is 1 / \nreg \is 1}   
&& w
}
\]
The induced closure function of the circuit satisfies the assertion graph.
The reason is that the two incoming edges at vertex $v$ disagree on the value
of node $\nregp$, therefore the value $\X$ is propagated to the outgoing edge of vertex $v$. 
The outgoing edge of vertex $v$ requires $\nreg$ to have value $1$,
therefore the consequent at that edge is satisfied. The
antecedent does not force any node take on value $\T$,
so the assertion graph is cautiously satisfied.  

But, the STE-assertion corresponding to the bottom initial path
\[
(\nregp \is 0) \en \N(\nreg \is 1) \Longrightarrow \N(\nreg \is 1)
\]
is not cautiously satisfied as every trajectory that satisfies the antecedent
gives node $\nreg$ value $\T$ at time 1.
\qed
\end{exa}

%

%% file: gste-mc.tex
\section{GSTE model checking}
\label{sec:mc}

In \cite{gsteIntroduction,gsteCaseStudy,abstractionInAction,gsteNotPublished}
model checking algorithms for GSTE are described. In this section, we show
the correspondence between the GSTE semantics presented in this paper and a
standard model checking algorithm. We do this by first relating our semantics
to a GSTE algorithm designed by ourselves, which uses a non-standard fixpoint
computation. We proceed by showing that our algorithm computes the same
result as the algorithm presented in \cite{gsteNotPublished}.

Furthermore, as we are concerned with precisely describing abstraction only,
we ignore extensions of GSTE algorithms such as backwards information flow
and fairness constraints.


%


\weg{
The soundness and completeness proofs are very much analog
to the technique used in STE. That is, first we define
the concepts of \emph{defining sequence graph} and
\emph{defining trajectory graph}. The defining sequence graph
of an antecedent (or consequent) is the weakest sequence graph
satisfying the antecedent (respectively consequent).
The defining trajectory is the weakest trajectory satisfying
the antecedent. Then, the \emph{fundamental theorem of GSTE}
states that in order to check that a GSTE assertions holds
we only need the check that the defining sequence graph of
the consequent is smaller than the defining sequence graph of
the antecedent.
}
\subsection{Fundamental theorem of GSTE}

Comparable to STE, GSTE model checking is based on the following:
Instead of checking that for every trajectory graph, the antecedent
implies the consequent, a unique weakest trajectory graph satisfying 
the antecedent is calculated. We call this graph the \emph{defining trajectory graph}. 

To check whether a circuit simply satisfies an assertion graph, 
it suffices to check whether the defining trajectory graph
satisfies the consequent part of the assertion graph.

Before giving the definition of the defining trajectory graph,
we first introduce the concept of the \emph{defining sequence graph}. 
The defining sequence graph of an antecedent  is the
unique weakest sequence satisfying the antecedent and is defined 
as follows.

Given an antecedent function $\ant : E \rightarrow \GTEL$,
and an assignment of symbolic constants $\phi$, 
we define  the \emph{defining sequence graph} of $\ant$ and $\phi$,
written $\dsf{\ant}$ by:
\[
\dsf{ant}(e) = \dsf{ant(e)}_{\textrm{state}} 
\]
where 
\[
\begin{array}{lcl}
\dsf{m \is b}_{\textrm{state}}(n)        & \;\; = \;\;  &  \twocases{b}{$ \text{if } m=n $}{\X}{$otherwise$}     \watruimte \\
\dsf{f_1 \en f_2}_{\textrm{state}}       &      =       & \dsf{f_1}_{\textrm{state}} \lub \dsf{f_2}_{\textrm{state}}                                    \watruimte \\
\dsf{P \rightarrow f}_{\textrm{state}}   &      =       & \twocases{\dsf{f}_{\textrm{state}}}{$ if $ \phi \models P }{\X}{$otherwise$} \watruimte \\
\end{array}
\]
\begin{prop}
$\dsf{\ant}$ is the weakest sequence graph satisfying $\ant$ and $\phi$.
\label{prop:defseqgraph}
\end{prop}
\proof
Trivial, by considering one edge at the time and induction on the structure
of the antecedent at that edge. \qed
%
Given an antecedent function $\ant : E \rightarrow \GTEL$,
a closure function $F$, and an assignment of symbolic constants $\phi$, 
we define  the \emph{defining trajectory graph} of $\ant$, $F$ and $\phi$,
written $\dtf{\ant}$ by:
\[
\dtf{\ant} = \Fsg (\dsf{ant})
\]
\begin{prop}
$\dtf{\ant}$ is the weakest trajectory graph satisfying $\ant$.
\label{prop:deftrajgraph}
\end{prop}

\proof
From $\Fsg$ being extensive, it follows directly that
$\dsf{ant} \leq \Fsg(\dsf{ant})$, so $\dtf{\ant} \models \ant$.

Suppose $\Tau$ is a trajectory graph satisfying $\ant$, then
$\dsf{ant} \leq T$. From monotonicity of $\Fsg$, it follows that
$\Fsg(\ant) \leq \Fsg(\Tau)$. As $\Tau$ is a fixpoint of $\Fsg$
it follows that $\Fsg(\ant) \leq \Tau$. \qed

\begin{theorem}[Fundamental Theorem of GSTE]
For each closure function $F$,
assignment of symbolic constants $\phi$,
and assertion graph $G=(V,E,\ant,\cons)$, 
\[
(\dsf{\cons} \leq \dtf{\ant})  \Leftrightarrow F \modS G
\] 
\end{theorem}

\proof
Directly from Proposition \ref{prop:deftrajgraph}. \qed
%
The fundamental theorem of GSTE states that 
to check whether a circuit with closure function $F$ 
satisfies an assertion graph, we only have to check
that, for each $\phi$, the defining trajectory graph of $F$
satisfies the consequent.
\subsection{GSTE Algorithm}
The GSTE algorithm 
calculates a symbolic representation 
of the defining trajectory graph of an antecedent. Then, it checks whether this
symbolic defining trajectory graph meets the consequent.


We first present a scalar version of the algorithm.

\subsubsection{A scalar GSTE-algorithm}
In (our version of the) scalar GSTE-algorithm, the defining trajectory graph of the antecedent
is calculated using the constructive version of Tarski's fixpoint theorem~\cite{tarski}.

\begin{prop}
For each $\Sigma$, 
the greatest fixpoint of the function $\FsgTwo$ is equal to limit $\Delta^{\Sigma}_*$ of the sequence
$\Delta^{\Sigma}_k=(\FsgTwo)^k(\T)$.
Here, $\T$ represents the sequence graph with the same edges and vertices as $\Sigma$ that gives
value $\T$ to each circuit node at each edge.
\label{prop:fp}
\end{prop}

\proof
A function is continuous if for all sequence $d_0,d_1,d_2,\ldots$ 
such that $d_{i+1} \leq d_i$ holds:
\[
f(\mathop{\lub}\limits_{k \in \Nat} d_k) = \mathop{\lub}\limits_{k \in \Nat} f(d_k)
\]
The constructive version of Tarski's fixpoint theorem~\cite{tarski} states that 
the greatest fixpoint of a monotone and continuous function $f$ on a complete
lattice is given by: $\mathop{\glb}\limits_{k \in \Nat} f^k(\T)$.

We will use this version of Tarski's fixpoint theorem to prove the proposition.
First, we prove that each monotonic function on a finite domain
is also continuous. Suppose $f$ is continuous on a finite domain, 
and $d_0,d_1,d_2,\ldots$ is a sequence such that $d_{i+1} \leq d_i$, 
then the sequence has a fixpoint $d_{*}$, thus:
\[
f(\mathop{\lub}\limits_{k \in \Nat} d_k) = f(d_*) 
\]
By monotonicity of $f$, also the sequence  $f(d_0),f(d_1),f(d_2),\ldots$
is increasing, so the sequence has the fix-point $f(d_*)$ as well. So:
\[
\mathop{\lub}\limits_{k \in \Nat} f(d_k) = f(d_*)
\]
Therefore, $\FsgTwo$ is both monotone and continuous. Thus:
\[
\gfp \Delta. \FsgTwo(\Delta) = \mathop{\glb}\limits_{k \in \Nat} (\FsgTwo)^k(\T) = \mathop{\glb}\limits_{k \in \Nat}
\Delta^{\Sigma}_k
\]
We prove by induction on $k$ that $\Delta^{\Sigma}_{k+1} \leq \Delta^{\Sigma}_k$ for each $k$.

The case for $k=0$ is trivial. The induction hypothesis is $\Delta_{k+1}^{\Sigma} \leq \Delta_k^{\Sigma}$.
We prove that $\Delta_{k+2}^{\Sigma} \leq \Delta_{k+1}^{\Sigma}$.
The case for $e$ is initial is trivial. Suppose $e$ is not initial. Then,
\[
\begin{array}{cll}
     & \Delta_{k+2}^{\Sigma}(e)          & \\
=    & \FsgTwo(\Delta_{k+1}^{\Sigma})(e) & (\text{Definition } \Delta_{k+2}^{\Sigma}) \\
=    & F(\; \Sigma(e) \tlub \mathop{\glb}\limits_{i \in \ine(e)} \nextf(\Delta_{k+1}^{\Sigma}(i)) \;) & (\text{Definition } \FsgTwo)     \\
\leq & F(\; \Sigma(e) \tlub \mathop{\glb}\limits_{i \in \ine(e)} \nextf(\Delta_k^{\Sigma}(i)) \;)   & (\text{Induction Hypothesis})     \\ 
=    & \FsgTwo(\Delta_{k+1}^{\Sigma})(e)                                                           & (\text{Defintion } \FsgTwo) \\
=    & \Delta_{k+1}^{\Sigma}(e)                                                                    & (\text{Definition } \Delta_{k+1}^{\Sigma}) \\
\end{array}
\]
So, the sequence $\Delta_0^{\Sigma},\Delta_1^{\Sigma},\Delta_2^{\Sigma}$ will eventually reach a fixpoint $\Delta_*^{\Sigma}$.

Thus:
$$
\gfp \Delta. \FsgTwo(\Delta) = \Delta^{\Sigma}_*\eqno{\qEd}
$$

\begin{defi}[Scalar GSTE-algorithm] 
Given an assertion graph $G$, and a closure function $F$, the scalar GSTE-algorithm
calculates for every $\phi$ the defining trajectory graph $\dtf{\ant}$ by calculating
$\Delta_{*}^{\dsf{\ant}}$ and checks whether 
\[
\dsf{\cons} \leq \dtf{\ant}
\]
If this check fails for any $\phi$ the algorithm returns False, otherwise
it returns True.
\end{defi}

\begin{prop}
The scalar algorithm is sound and complete with respect to the presented
semantics for GSTE.
\end{prop}

\proof
Directly from the fundamental theorem of GSTE and Proposition \ref{prop:fp}. \qed

\head{Comparing with earlier presentation}
In \cite{gsteNotPublished} the fixpoint is calculated in a slightly
different way. If we adjust the presentation to use the closure function
$F$ instead of a transition relation, 
the following sequence is defined for a given sequence graph $\Sigma$:
\[
\begin{array}{lcl}
 \Gamma^{\Sigma}_0(e)      & = & \twocases{F(\Sigma(e))}{e \text{ is an initial edge}}{\T}{\text{otherwise}} \\
 \Gamma^{\Sigma}_{k+1}(e)  & = & \Gamma^{\Sigma}_{k}(e)  \tglb  
                                    F(\; \Sigma(e) \tlub \mathop{\glb}\limits_{i \in \ine(e)} \nextf(\Delta_k(i)) \;)
\end{array}
\]

\begin{prop} For each $\Sigma$, 
$
\Gamma^{\Sigma}_{*} = \Delta^{\Sigma}_{*}
$.
\end{prop}

\proof
By definition,
\[
\begin{array}{lcl}
\Delta^{\Sigma}_{0}(e)(n) & = & \T \\ 
\Delta^{\Sigma}_{k+1}(e)  & = & 
\twocases
{F(\Sigma(e))}{\text{$e$ is initial}}
{F(\; \Sigma(e) \tlub \mathop{\glb}\limits_{i \in \ine(e)} \nextf(\Delta_k(i)) \;)}{\text{otherwise}} 
\end{array}
\]
We prove by induction on $k$ that for each initial edge $e$, 
for each $k$, $\Gamma^{\Sigma}_k= F(\Sigma(e))$. The base case
is trivial. Now suppose for each initial edge $e$, $\Gamma^{\Sigma}_k(e)= F(\Sigma(e))$, then
for an arbitrary initial edge $e$:
\[
\begin{array}{cll}
   & \Gamma^{\Sigma}_{k+1}(e) \\ 
 = &  \Gamma^{\Sigma}_{k}(e) \tglb F(\; \Sigma(e) \tlub \ldots \;) 
         & (\text{Definition } \Gamma^{\Sigma}_{k+1}) \\
 = & F(\Sigma(e)) \tglb F(\; \Sigma(e) \tlub \ldots \;) 
         & (\text{Induction Hypothesis}) \\
 = & F(\Sigma(e)) & (\text{Property } \lub,\glb) 	 
\end{array}
\]
So, for each initial edge and $k>0$, $\Gamma^{\Sigma}_{k}(e) = \Delta^{\Sigma}_{k}(e)$.

We prove by induction on $k$ that for each non-initial edge $e$, 
for each $k>0$, 
\[
\Gamma^{\Sigma}_{k}(e) = \Delta^{\Sigma}_{k}(e) 
\]
In the base-case, $k$ is equal to $1$,
\[
\begin{array}{cll}
     & \Gamma^{\Sigma}_1(e)     & \\
=    & \Gamma^{\Sigma}_{0}(e)  \tglb  
       F(\; \Sigma(e) \tlub \mathop{\glb}\limits_{i \in \ine(e)} \nextf(\Gamma_0(i)) & (\text{Definition } \Gamma^{\Sigma}_{k+1})\\ 
=    & \T  \tglb  
       F(\; \Sigma(e) \tlub \mathop{\glb}\limits_{i \in \ine(e)} \nextf(\T(i)) & (\text{Definition } \Gamma^{\Sigma}_{0})\\ 
=    & F(\; \Sigma(e) \tlub \mathop{\glb}\limits_{i \in \ine(e)} \nextf(\T(i)) & (\text{Property } \glb)\\ 
=    & F(\; \Sigma(e) \tlub \mathop{\glb}\limits_{i \in \ine(e)} \nextf(\Delta_0(i)) & (\text{Definition } \Delta^{\Sigma}_{0})\\ 
=    & \Delta^{\Sigma}_{1} & (\text{Definition } \Delta^{\Sigma}_{1})\\ 
\end{array}
\]
Now suppose $\Gamma^{\Sigma}_{k}(e) = \Delta^{\Sigma}_{k}(e) $,
then:
\[
\begin{array}{cll}
     & \Gamma^{\Sigma}_{k+1}(e)     & \\
=    & \Gamma^{\Sigma}_{k}(e)  \tglb  
       F(\; \Sigma(e) \tlub \mathop{\glb}\limits_{i \in \ine(e)} \nextf(\Gamma_{k}(i)) & (\text{Definition } \Gamma^{\Sigma}_{k+2})\\ 
=    & \Delta^{\Sigma}_{k}(e)  \tglb F(\; \Sigma(e) \tlub \mathop{\glb}\limits_{i
\in \ine(e)} \nextf(\Delta_{k}(i)) 
             & (\text{Induction Hypothesis}) \\
=    & \Delta^{\Sigma}_{k}(e) \glb \Delta^{\Sigma}_{k+1}(e) & (\text{Definition } \Delta^{\Sigma}_{k+1}(e)) \\
=    & \Delta^{\Sigma}_{k+1}(e) & (\text{Property } \glb, \Delta^{\Sigma}_{k+1}(e) \leq \Delta^{\Sigma}_{k}(e)) 
\end{array}
\]
So, for each non-initial edge and $k>0$, 
$\Gamma^{\Sigma}_{k}(e) = \Delta^{\Sigma}_{k}(e) $. So, for $k>0$, $\Gamma^{\Sigma}_{k} = \Delta^{\Sigma}_{k}$.
Thus, $\Gamma^{\Sigma}_{*} = \Delta^{\Sigma}_{*}$.  \qed

\subsubsection{A symbolic GSTE-algorithm} In actual implementations of GSTE,
the above algorithm is implemented symbolically. That is, instead of calculating
the defining trajectory graph for a specific valuation $\phi$, it calculates,
using BDDs, a symbolic defining trajectory graph in terms of the
symbolic constants in $\phi$. 

Then, a BDD is constructed that specifies under which conditions on the symbolic
constants the symbolic defining trajectory graph satisfies the consequent. If this
BDD is equal to the logical constant True, the property is proven. Otherwise, the BDD indicates
for which valuations of the symbolic constants the antecedent does not imply 
the consequent.

%% file: future.tex
\section{Future Work}

\subsubsection*{Extension to the semantics for GSTE}

There exist several extensions of the GSTE 
algorithm that considerably improve the algorithm's proving power.
Examples of such extensions are \emph{precise
nodes} \cite{abstractionInAction,gsteCaseStudy} and \emph{knots} \cite{monitor2}.
We would like to give semantic characterisations of these
extensions. 

In \cite{gsteNotPublished,gsteIntroduction,abstractionInAction}, a backwards algorithm for GSTE is described.
Using this algorithm properties can be proven that depend
on a backwards (that is, from outputs to inputs, and from time
$t+1$ to $t$) information flow. In \cite{gsteNotPublished,abstractionInAction} a semantics for
this form of GSTE is given. The semantics is
however not faithful as the algorithm is incomplete w.r.t. the semantics~\cite{gsteNotPublished}.
A faithful semantics for this form of GSTE  could be a topic of future work.  

\subsubsection*{SAT-based GSTE model checking}

The model checking algorithms for GSTE described in the current literature
are based on BDDs. 
In previous work we described how a faithful semantics for STE~\cite{steFaithful}
enabled us to construct a new SAT-based model checking algorithm
for STE~\cite{roorda}.
In the same way, our faithful semantics for GSTE could be used to 
construct a SAT-based model checking algorithm for GSTE. The aim would be to
create a tool very much like \emph{satGSTE} \cite{GSTESAT} that actually
respects the GSTE semantics, so that it can possibly find {\em all} counter
examples. In this way, the tool could be used seamlessly  in conjunction with a GSTE
model checker.


\subsubsection*{Monitor circuits for GSTE assertion graphs}


In \cite{monitor1,monitor2} methods for automatic
construction of monitor circuits for GSTE assertion graphs are described.
%

The papers explain how monitor circuits
can be used
to quickly debug and refine GSTE specifications before trying to use,
more labour intensive, GSTE model checking. 

The monitor circuits implement the $\forall$-semantics for GSTE.
However, as explained in this paper, the GSTE model checking algorithms are not faithful
to this semantics. Therefore, monitor circuits cannot be used
to debug and refine assertion graphs that are true in the $\forall$-semantics,
but yield a spurious counter-example when trying to prove them with a GSTE model checker.
%
Future work could consist of constructing monitor circuits that
can be used to debug and refine assertion graphs in this class.
Here, the faithful semantics for GSTE can be used as a starting
point.

\subsubsection*{Reasoning about GSTE assertion graphs}

Using the construction of monitor circuits for GSTE assertion
graphs, \cite{reasoning} describes two algorithms that can be
used in compositional verification using GSTE. 
The first algorithm decides whether one assertion graph
implies another.
The second algorithm can be used to model check an assertion
graph under the assumption that another assertion graph is true.

The algorithms, and the corresponding soundness and completeness proofs,
are based on the $\forall$-semantics. Therefore, as the algorithms are based
on GSTE model checking, the methods are incomplete when abstraction
is used. A possible direction for future work is explaining how the GSTE abstraction
affects the completeness of the algorithms.

%% file: conclusion.tex
\section{Conclusion}

The semantics for GSTE given in \cite{gsteNotPublished,abstractionInAction} are not faithful to the proving power of
GSTE model checking algorithms, that is, the algorithms are incomplete with respect to the
semantics. The reason is that the the semantics do not capture the
abstraction used in GSTE precisely. 

The abstraction used in GSTE makes it hard to understand why a specific property can, or
cannot, be proven by GSTE. The  semantics mentioned above cannot help the user in doing so.
So, in the current situation, users of GSTE often have to revert to the GSTE algorithm to
understand why a property can or cannot be proven by GSTE.

In this paper, we have presented a semantics for GSTE that is faithful to the proving power of
the main GSTE model checking algorithm.
We believe that this semantics is an important contribution to 
the research on GSTE for at least two reasons.

First of all, a faithful semantics makes GSTE more accessible to novice users: 
a faithful semantics enables users to understand the abstraction used in GSTE,
without having to understand the details of the model checking algorithm.

Furthermore, 
a faithful semantics for GSTE can be used
as basis for research on new GSTE model checking algorithms and other GSTE tools.
To illustrate this, in previous work~\cite{roorda}, we described a new SAT-based model checking algorithm
for STE and proven that it is sound and complete w.r.t. to our faithful semantics
for STE presented in~\cite{steFaithful}. Without a faithful semantics for STE, we would have been forced to prove the
correctness of our algorithm by relating it to other model checking algorithms
for STE. This is clearly a more involved and less elegant approach. 
In fact, we believe that without constructing a faithful semantics for STE first,
we would not have obtained the level of understanding of STE needed to develop the
new SAT-based model checking algorithm. 

In the same way, we expect that the faithful semantics for GSTE presented
in this paper will open the door for new research on GSTE model checking algorithms and other GSTE tools.
%


%

\subsubsection*{Acknowledgements}

Thanks to Tom Melham, Mary Sheeran, Rachel Tzoref, and the anonymous referees
for commenting on earlier drafts of this paper.

%% file: main.bbl
\begin{thebibliography}{10}

\bibitem{methodology}
Mark Aagaard, Robert~B. Jones, Thomas~F. Melham, John~W. O'Leary, and
  Carl-Johan~H. Seger.
\newblock A methodology for large-scale hardware verification.
\newblock In Warren A.~Hunt Jr. and Steven~D. Johnson, editors, {\em FMCAD},
  volume 1954 of {\em Lecture Notes in Computer Science}, pages 263--282.
  Springer, 2000.

\bibitem{DBLP:conf/fmcad/2002}
Mark Aagaard and John~W. O'Leary, editors.
\newblock {\em Formal Methods in Computer-Aided Design, 4th International
  Conference, FMCAD 2002, Portland, OR, USA, November 6-8, 2002, Proceedings},
  volume 2517 of {\em Lecture Notes in Computer Science}. Springer, 2002.

\bibitem{forte}
{FORTE}.
\newblock \\
  \textsf{http://www.intel.com/software/products/opensource/tools1/verificatio%
n}.

\bibitem{monitor1}
Alan~J. Hu, Jeremy Casas, and Jin Yang.
\newblock Efficient generation of monitor circuits for {GSTE} assertion graphs.
\newblock In {\em International Conference on Computer-Aided Design (ICCAD)},
  pages 154--160. IEEE Computer Society / ACM, 2003.

\bibitem{reasoning}
Alan~J. Hu, Jeremy Casas, and Jin Yang.
\newblock Reasoning about {GSTE} assertion graphs.
\newblock In Daniel Geist and Enrico Tronci, editors, {\em Correct Hardware
  Design and Verification Methods (CHARME)}, volume 2860 of {\em Lecture Notes
  in Computer Science}, pages 170--184. Springer, 2003.

\bibitem{indexing}
Thomas~F. Melham and Robert~B. Jones.
\newblock Abstraction by symbolic indexing transformations.
\newblock In Aagaard and O'Leary \cite{DBLP:conf/fmcad/2002}, pages 1--18.

\bibitem{monitor2}
Kelvin Ng, Alan~J. Hu, and Jin Yang.
\newblock Generating monitor circuits for simulation-friendly {GSTE} assertion
  graphs.
\newblock In {\em International Conference on Computer Design (ICCD)}, pages
  409--416. IEEE Computer Society, 2004.

\bibitem{roorda}
Jan-Willem Roorda and Koen Claessen.
\newblock A new {SAT}-based algorithm for symbolic trajectory evaluation.
\newblock In Dominique Borrione and Wolfgang~J. Paul, editors, {\em Correct
  Hardware Design and Verification Methods (CHARME)}, volume 3725 of {\em
  Lecture Notes in Computer Science}, pages 238--253. Springer, 2005.

\bibitem{steFaithful}
Jan-Willem Roorda and Koen Claessen.
\newblock Explaining symbolic trajectory evaluation by giving it a faithful
  semantics.
\newblock In Dima Grigoriev, John Harrison, and Edward~A. Hirsch, editors, {\em
  International Computer Science Symposium in Russia (CSR)}, volume 3967 of
  {\em Lecture Notes in Computer Science}, pages 555--566. Springer, 2006.

\bibitem{highlevel}
Thomas Schubert.
\newblock High level formal verification of next-generation microprocessors.
\newblock In {\em Design Automation Conference (DAC)}, pages 1--6. ACM, 2003.

\bibitem{vardi}
Roberto Sebastiani, Eli Singerman, Stefano Tonetta, and Moshe~Y. Vardi.
\newblock {GSTE} is partitioned model checking.
\newblock In Rajeev Alur and Doron Peled, editors, {\em Computer-Aided
  Verification (CAV)}, volume 3114 of {\em Lecture Notes in Computer Science},
  pages 229--241. Springer, 2004.

\bibitem{ste}
Carl-Johan~H. Seger and Randal~E. Bryant.
\newblock Formal verification by symbolic evaluation of partially-ordered
  trajectories.
\newblock {\em Formal Methods in System Design}, 6(2), 1995.

\bibitem{tarski}
A.~Tarski.
\newblock A lattice theoretical fixpoint theorem and its applications.
\newblock {\em Pacific J. of Mathematics}, 5:285--309, 1955.

\bibitem{GSTESAT}
Jin Yang, Rami Gil, and Eli Singerman.
\newblock {satGSTE}: Combining the abstraction of {GSTE} with the capacity of a
  {SAT} solver.
\newblock In {\em Designing Correct Circuits ({DCC}), A satellite event of the
  ETAPS 2004 group of conferences}, 2004.

\bibitem{gsteCaseStudy}
Jin Yang and Amit Goel.
\newblock {GSTE through a case study.}
\newblock In Lawrence~T. Pileggi and Andreas Kuehlmann, editors, {\em
  International Conference on Computer-Aided Design (ICCAD)}, pages 534--541.
  ACM, 2002.

\bibitem{gsteIntroduction}
Jin Yang and C.-J.~H. Seger.
\newblock Introduction to generalized symbolic trajectory evaluation.
\newblock In {\em International Conference on Computer Design (ICCD)}, pages
  360--367, Washington - Brussels - Tokyo, 2001. IEEE.

\bibitem{gsteNotPublished}
Jin Yang and Carl Seger.
\newblock Generalized symbolic trajectory evaluation.
\newblock Unpublished draft, 2001.

\bibitem{abstractionInAction}
Jin Yang and Carl-Johan~H. Seger.
\newblock Generalized symbolic trajectory evaluation - abstraction in action.
\newblock In Aagaard and O'Leary \cite{DBLP:conf/fmcad/2002}, pages 70--87.

\end{thebibliography}
